\documentclass[twocolumn,english,aps,pra,eqsecnum]{revtex4-1}
\usepackage[T1]{fontenc}
\usepackage[latin9]{inputenc}
\usepackage{color}
\usepackage{amsmath}
\usepackage{amssymb}
\usepackage{graphicx}
\usepackage{esint}

\makeatletter

\newcommand*\LyXThinSpace{\,\hspace{0pt}}
\providecommand{\tabularnewline}{\\}

\usepackage[english]{babel}

\usepackage{babel}

\makeatother

\usepackage{babel}
\begin{document}

\title{Einstein-Podolsky-Rosen steering, depth of steering and planar spin
squeezing in two-mode Bose-Einstein condensates }

\author{Laura Rosales-Zárate$^{1}$, B. J. Dalton$^{2}$ and M. D. Reid$^{2}$}

\affiliation{$^{1}$Centro de Investigaciones en Óptica A.C., León, Guanajuato
37150, México ~~\\
 $^{2}$Centre for Quantum and Optical Science, Swinburne University
of Technology, Melbourne 3122, Australia~~\\
 }
\begin{abstract}
We show how one can prepare and detect entanglement and Einstein-Podolsky-Rosen
(EPR) steering between two distinguishable groups (modes) of atoms
in a Bose-Einstein condensate (BEC) atom interferometer. Our paper
extends previous work that developed criteria for two-mode entanglement
and EPR steering based on the reduced variances of two spins defined
in a plane. Observation of planar spin squeezing will imply entanglement,
and sufficient planar spin squeezing implies EPR steering, between
the two groups of atoms. By using a two-mode dynamical model to describe
BEC interferometry experiments, we show that the two-mode entanglement
and EPR steering criteria are predicted to be satisfied for realistic
parameters. The reported observation of spin squeezing in these parameter
regimes suggests it is very likely that the criteria can be used to
infer an EPR steering between mesoscopic groups of atom\textcolor{black}{s,
provided the total atom number can be determined to sub-Poissonian
uncertainty. The criteria also apply to a photonic Mach-Zehnder interferometer.
F}inally, we give a method based on the amount of planar spin squeezing
to determine a lower bound on the number of particles that are genuinely
comprise the two-mode EPR steerable state $-$ the so-called two-mode
EPR steering depth. 
\end{abstract}
\maketitle

\section{Introduction}

The detection of entanglement between mesoscopic groups of atoms is
an important milestone. Two systems are entangled if the overall wavefunction
cannot be factorised into parts associated solely with each system.
While there has been significant progress in entangling microscopic
systems \cite{bellexp-1}, it is the entanglement of macroscopic massive
systems that provides some of the strangest predictions of quantum
mechanics \cite{Schrodinger-1-2}.\textcolor{black}{{} }This has
motivated experiments that report entanglement and quantum correlations
for massive systems, such as thermal atomic ensembles,\textcolor{black}{{}
cooled atoms, }and Bose-Einstein condensates (BEC) \cite{esteve-2,GrossTutorial-1,Gross2010-1,herald3000-1,Philipp2010-1,eprenthiedel-1,treutlein-exp-bell,sciencepairsent,mitchell,neweprbec,epranubec-1,jo,eprnaturecommun,milestonewhy-1,bell-kasevich}.
Very recently, entanglement has been detected between spatially separated
clouds formed from a BEC \cite{EntAtoms,SteerAt-obert,treu-matteo}.

A subtlety exists with the interpretation of multi-atom experiments:
detecting entanglement within an atomic group (or between two groups)
does not strictly imply that more than two atoms are entangled. In
light of this, efforts have been made to calibrate the number of atoms
that genuinely comprise the entangled state, the so-called ``depth
of entanglement'' \cite{sm-2,toth-planar-ent}. This has led to experimental
evidence for large numbers of atoms genuinely entangled at one location
\cite{Gross2010-1,Philipp2010-1}. However, so far, the methods of
calibration have mainly focused on the entanglement between particles
that are in principle distinguishable \cite{sorzollcirac-1,sm-2}.
This contrasts with the notion of the ``depth of the entanglement''
between two groups of indistinguishable bosonic atoms, such as occurs
for a Bose-Einstein condensate.

The detection of mesoscopically entangled atomic states also leads
to the question: what type of entanglement is certified? A subset
of entangled states gives rise to nonlocal effects, such as the Einstein-Podolsky-Rosen
(EPR) paradox and failure of local hidden variable theories \cite{eprbell-1,bell-1,Schrodinger-1-2,schrod-steering}.
\emph{EPR steerable} states are generalisations of the states considered
by EPR in their 1935 paradox, which reveal an inconsistency between
local realism and the completeness of quantum mechanics \cite{hw-1-1,schrod-steering,eprbell-1,eprmrrmp,eric-steer}.
EPR steerable states are important from a fundamental perspective
and also have applications for quantum information processing \cite{collapsewave-1,steerapp-1}.
EPR steering is required for Bell's form of nonlocality, which leads
to a falsification of all local hidden variable theories \cite{hw-1-1}.

There has been a growing experimental interest in EPR steering correlations
for atoms. Collective measurements have been used to indicate the
presence of Bell correlations (and hence EPR steering) within a BEC
or thermal ensemble of atoms \cite{treutlein-exp-bell,acinnonlocal-1,bell-kasevich}.
Experiments have reported observation of entanglement and EPR steering
correlations between distinguishable atomic groups \cite{SteerAt-obert,treu-matteo,eprnaturecommun,eprenthiedel-1,Gross2010-1,EntAtoms}.
The issue of whether entanglement occurs between particles or modes
for identical particle systems has become topical, and has been analysed
in some recent theoretical papers \cite{bryan-reviews,PlenioInd}
(see also Appendix D herein). There has however, to our knowledge,
been as of yet no quantification given of the number of atoms genuinely
involved in an EPR steerable state.

In this paper, we introduce the concept of ``\emph{depth of EPR steering}''.
We derive criteria to give evidence of two-mode EPR steerable states
genuinely comprised of many atoms, and further show how such steerable
states are predicted to be created in a two-mode BEC interferometer.
Methods to generate entangled and steering correlations have been
proposed based on four- or two-component BECs using either dynamical
evolution or cooling to a ground state\textcolor{red}{{} }\textcolor{black}{\cite{hesteer-1,asineprbec,bogdaneprbec,bryanlibby-1}.
}Recent EPR steering experiments exploit four-component BECs to generate
the correlations \cite{SteerAt-obert,treu-matteo}. Here, we provide
a different approach, based on the dynamical evolution of a two-component
BEC.

Entanglement and EPR-steering between two modes can be inferred from
the observation of planar quantum spin squeezing (PQS). The criteria
of this paper are based on the sum of two spin variances, as given
by\textcolor{red}{{} }the Hillery-Zubairy parameter \cite{hillzub-1}
\begin{equation}
E_{HZ}\equiv\frac{(\Delta\hat{S}_{x})^{2}+(\Delta\hat{S}_{y})^{2}}{\langle\hat{N}\rangle/2}\label{eq:ehz}
\end{equation}
Using the Schwinger representation, the spins are associated with
two modes. Denoting the boson annihilation operators for each mode
by $\hat{a}$ and $\hat{b}$, the total number operator $\hat{N}$
and spin $S$ are $\hat{N}=\hat{a}^{\dagger}\hat{a}+\hat{b}^{\dagger}\hat{b}$
and $S=\langle\hat{N}\rangle/2$. Hillery and Zubairy showed that
a sufficient condition for entanglement between the two modes is $E_{HZ}<1$
\cite{hillzub-1}. On the other hand, the similar condition for EPR-steering
is $E_{HZ}<0.5$ \cite{hesteer-1,bryan-epr-steer}\textcolor{black}{.
Planar quantum spin squeezing occurs when the noise in both spins
is sufficiently reduced, so that the sum of the variances is below
the shot noise level \cite{cj-2,toth-planar-ent}. For a single spin,
spin squeezing is achieved when $(\Delta\hat{S}_{y})^{2}<|\langle\hat{S}_{x}\rangle|/2$
\cite{ueda-1}. When the magnitude of the spin $S_{x}$ is maximised,
so that $\langle\hat{S}_{x}\rangle=\langle N\rangle/2$, this corresponds
to a variance below the shot noise level, $(\Delta\hat{S}_{y})^{2}<S/2$.
PQS occurs for $(\Delta\hat{S}_{x})^{2}+(\Delta\hat{S}_{y})^{2}<|\langle S_{||}\rangle|$
where $|\langle S_{||}\rangle|=\sqrt{\langle\hat{S}_{x}\rangle^{2}+\langle\hat{S}_{y}\rangle^{2}}$
\cite{toth-planar-ent}, which when $|\langle S_{||}\rangle|$ is
maximised at $\langle\hat{N}\rangle/2$ corresponds to $E_{HZ}<1$.}

\textcolor{black}{In this paper, we show that EPR steering correlations
can be certified using the Hillery-Zubairy parameter, and that a method
similar to that developed by Sørensen and Mølmer \cite{sm-2} can
be used to calibrate the number of atoms in the steerable state. Our
calibration of a lower bound on how many atoms are involved in the
two-mode steerable state is based on the tight value $C_{S}$ for
the minimum of the sum of the planar spin variances ($(\Delta\hat{S}_{x})^{2}+(\Delta\hat{S}_{y})^{2}\geq C_{S}$)
given a fixed spin $S$ value, as derived by He et al \cite{cj-2}.
We also explain how the sensitivity of the estimate might be improved,
if $\langle S_{||}\rangle$ is also measured, based on the lower bound
of the functions $(\Delta\hat{S}_{x})^{2}+(\Delta\hat{S}_{y})^{2}$
for a given $S$ and $\langle S_{||}\rangle$, recently derived by
Vitagliano et al \cite{toth-planar-ent}. }Although the $E_{HZ}$
signature involves collective spin measurements, thereby not directly
testing nonlocality, we note that the criteria can be rewritten in
terms of quadrature phase amplitudes to give a method that allows
local measurements on individual subsystems \cite{eprenthiedel-1}.

By analysing the predictions for a simple two-mode BEC interferometer
in the limit of stationary wavefunctions, we follow Li et al \cite{yun-li,yunli-2}
to show that spin squeezing of the spin vector $\hat{S}_{\theta}$
in the $yz$ plane is possible for certain $\theta$. We then show
that this implies entanglement between suitably rotated modes that
can be created in the interferometer using the atom-optics equivalent
of phase shifts and beam splitters. In fact, entanglement can be created
without the BEC nonlinearity \cite{hesteer-1}. However, the nonlinearity
is required to create sufficient spin squeezing to allow detection
of steering via the Hillary-Zubairy parameter. A spin squeezing of
$S_{\theta}$ has been observed in the experiments of Riedel et al
\cite{Philipp2010-1}, which suggests that the observation of EPR
steering is also possible, provided one can also detect the predicted
reduction in the variance of the spin $S_{x}$ which describes the
Bloch vector. This requires control of the number fluctuations of
the total atom number $\hat{N}$.

In the conclusion, we discuss the effect of the \textcolor{black}{dynamical
spatial variation} of the wavefunction, as given in Li et al \cite{yunli-2}
and accounted for in the multi-mode models of a BEC interferometer
by Opanchuk et al \cite{bogdanepl,Egorov-1}. The atom interferometer
is realisable in different forms including where the modes are associated
with two hyperfine atomic levels confined to the potential wells of
an optical lattice \cite{esteve-2,Gross2010-1}; are the outputs of
a BEC beam splitter on an atom chip \cite{Philipp2010-1,Egorov-1};
and where large numbers of atoms and/ or spatial separations are possible
\cite{anu-exp-1,stanford-exp-1,Egorov-1}. Planar spin squeezing has
been observed for thermal atomic ensembles with significant applications
\cite{mitchell-newj-ourn,timpaper-2,pssexp}. The methods of this
paper can also be applied to optical experiments based on polarisation
squeezing \cite{pol-sq}.

\section{Criteria for EPR steering and entanglement}

\subsection{Entanglement and EPR steering }

Consider two systems $A$ and $B$ described by a quantum density
operator $\rho$. Assuming each system is a single mode, we define
the boson creation and destruction operators $\hat{a}^{\dagger}$,
$\hat{a}$, $\hat{b}^{\dagger}$, $\hat{b}$ for $A$ and $B$ respectively.
The two systems are said to be entangled if the combined system cannot
be described by a separable density operator 
\begin{equation}
\rho=\sum_{R}P_{R}\rho_{A}^{R}\rho_{B}^{R}\label{eq:sep}
\end{equation}
\cite{wernerperes-2}. In this notation, $\rho_{A}^{R}$ and $\rho_{B}^{R}$
are density operators for systems $A$ and $B$ respectively, and
$P_{R}$ are probabilities satisfying $\sum_{R}P_{R}=1$ and $P_{R}>0$.
Where the systems $A$ and $B$ are spatially separated, the entangled
state can give rise to nonlocality \cite{bell-1,eprbell-1}. EPR steering
of $B$ by $A$ is certified if there is a failure of all local hidden
state (LHS) models, where the averages for locally measured observables
$\hat{X}_{A}$ and $\hat{X}_{B}$ are given as \cite{hw-1-1} 
\begin{equation}
\langle\hat{X}_{B}\hat{X}_{A}\rangle=\int_{\lambda}P(\lambda)d\lambda\langle\hat{X}_{B}\rangle_{\rho,\lambda}\langle\hat{X}_{A}\rangle_{\lambda}\label{eq:lhs}
\end{equation}
The states symbolised by $\lambda$ are the hidden variable states
introduced in Bell's local hidden variable theories. Here, $P(\lambda)$
is the the probability density satisfying $\int_{\lambda}P(\lambda)d\lambda=1$
\cite{bell-1} and $\langle X_{A}\rangle_{\lambda}$ is the average
of $\hat{X}_{A}$ given the system is in the hidden state $\lambda$.
To test for steering, an additional constraint has been introduced.
This is symbolised by the subscript $\rho$ for the averages calculated
for $B$. The average $\langle\hat{X}_{B}\rangle_{\rho,\lambda}$
is constrained to be consistent with that of a local quantum density
operator $\rho_{B}^{\lambda}$. The states can be steerable ``one-way''
($B$ by $A$) as evidenced by failure of the above model (\ref{eq:lhs}).
Alternatively, by exchanging $A\longleftrightarrow B$ in the model,
failure of the LHS model 
\begin{equation}
\langle\hat{X}_{B}\hat{X}_{A}\rangle=\int_{\lambda}P(\lambda)d\lambda\langle\hat{X}_{A}\rangle_{\rho,\lambda}\langle\hat{X}_{B}\rangle_{\lambda}\label{eq:lhs-1}
\end{equation}
implies steering of $A$ by $B$. It is also possible to demonstrate
steering ``two-ways'' ($B$ by $A$, and $A$ by $B$) \cite{oneway-1}.

\subsection{Criterion for EPR steering based on spin variances}

All separable models (\ref{eq:sep}) imply the Hillery-Zubairy inequality
\cite{hillzub-1} 
\begin{equation}
|\langle\hat{a}^{\dagger}\hat{b}\rangle|^{2}\le\langle\hat{a}^{\dagger}\hat{a}\hat{b}^{\dagger}\hat{b}\rangle\label{eq:hzab}
\end{equation}
The LHS model (\ref{eq:lhs}) implies the inequality 
\begin{equation}
|\langle\hat{a}^{\dagger}\hat{b}\rangle|^{2}\le\langle\hat{a}^{\dagger}\hat{a}\hat{b}^{\dagger}\hat{b}\rangle+\langle\hat{b}^{\dagger}\hat{b}\rangle/2\label{eq:hzsteerab}
\end{equation}
derived by Cavalcanti et al \cite{cavalunified-1}. These inequalities
if violated confirm entanglement and EPR-steering (of $B$ by $A$)
respectively. The inequalities can be expressed in terms of Schwinger
spin observables $\hat{S}_{z}=(\hat{a}^{\dagger}\hat{a}-\hat{b}^{\dagger}\hat{b})/2$,
$\hat{S}_{x}=(\hat{a}^{\dagger}\hat{b}+\hat{a}\hat{b}^{\dagger})/2$,
$\hat{S}_{y}=(\hat{a}^{\dagger}\hat{b}-\hat{a}\hat{b}^{\dagger})/2i$
and $\hat{N}=\hat{a}^{\dagger}\hat{a}+\hat{b}^{\dagger}\hat{b}$ ($\hbar=1$)
to give the conditions 
\begin{equation}
E_{HZ}<1\label{eq:ehzent}
\end{equation}
and \textcolor{black}{{} 
\begin{equation}
E_{HZ}<\frac{\langle\hat{a}^{\dagger}\hat{a}\rangle}{\langle\hat{N}\rangle}\label{eq:ehzsteer}
\end{equation}
}sufficient to certify entanglement \cite{hillzub-1} and\textcolor{black}{{}
EPR-steering ($B$ by $A$) respectively \cite{bryan-epr-steer}.
States that are not steerable will satisfy both LHS models (\ref{eq:lhs})
and (\ref{eq:lhs-1}), defined to test steering of system $A$ or
steering of system $B$. Hence non-steerable states satisfy both }$E_{HZ}\geq\frac{\langle\hat{a}^{\dagger}\hat{a}\rangle}{\langle\hat{N}\rangle}$
and $E_{HZ}\geq\frac{\langle\hat{b}^{\dagger}\hat{b}\rangle}{\langle\hat{N}\rangle}$,
which implies that $E_{HZ}\geq0.5$. Thus, the condition 
\begin{equation}
E_{HZ}<0.5\label{eq:half-steer}
\end{equation}
will imply EPR steering \cite{hesteer-1,bryan-epr-steer}. \textcolor{black}{{} }

\section{depth of two-mode entanglement and EPR steering }

In this section, we show that the degree of reduction in the value
of $E_{HZ}$ will place a lower bound on the minimum number of bosons
in the two-mode entangled or EPR steerable state. For a system of
fixed spin $S$, He et al determine the bounds $C_{S}$ of the quantum
uncertainty relation (for $S\neq0$) \cite{cj-2}: $(\Delta\hat{S}_{x})^{2}+(\Delta\hat{S}_{y})^{2}\geq C_{S}$.
We normalise this expression, to write

\begin{equation}
\frac{(\Delta\hat{S}_{x})^{2}+(\Delta\hat{S}_{y})^{2}}{S}\geq\frac{C_{S}}{S}\equiv\tilde{C_{S}}\label{eq:CJUR-1}
\end{equation}
The $\tilde{C}_{S}$ is a coefficient that determines the tight minimum
value of the sums of the two variances: The values are found in \cite{cj-2}
and are plotted in Figure 1. \textcolor{black}{He et al compute $C_{S}$
using a numerical optimisation procedure. The lower bounds for $S=1/2$
and $S=1$ were derived in Ref. \cite{hofmann2}. He et al also determine
a precise asymptotic dependence $C_{S}\sim a_{0}S^{2/3}$ for large
$S$ by analytic means. The relation indicates the amount of noise
reduction that is possible in just two spin components and has been
used for the derivation of entanglement criteria \cite{cj-2}, interferometry
and phase estimation \cite{timpaper-2}, and for placing ultimate
constraints on levels of planar spin squeezing \cite{pssexp,mitchell-newj-ourn}. }

\begin{figure}
\begin{centering}
\includegraphics[width=0.6\columnwidth]{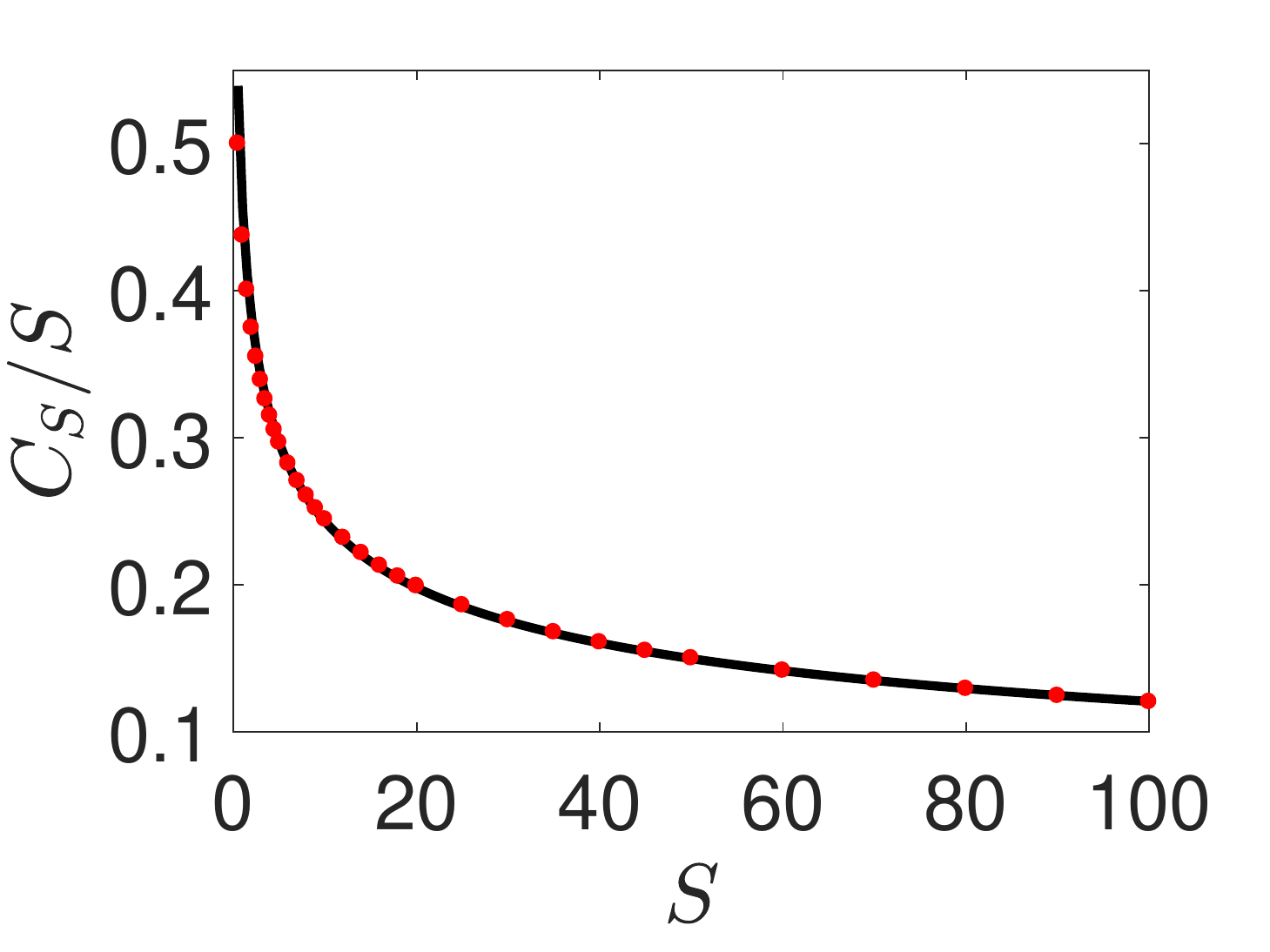} 
\par\end{centering}
\caption{Values of $\tilde{C}_{S}$ for a system of fixed spin $S$. Results
are taken from He et al \cite{cj-2}. \label{fig:system-two modes}\textcolor{red}{{} }}
\end{figure}

\subsection{Depth of two-mode entanglement}

The curves of Figure 1 can be used in a similar way to the Sørensen-Mølmer
curves \cite{sm-2} to determine a lower bound on the number of boson
particles that \emph{genuinely} comprise a pure two-mode entangled
state. This we refer to as the ``\emph{two-mode entanglement depth}''.
We note that the number of particles in the entangled state is \emph{not}
simply given by the mean $\langle\hat{N}\rangle$, because in general
an experimental system will be a mixture of pure states. It is therefore
possible that mixed entangled states with large $\langle\hat{N}\rangle$
arise from highly populated separable states. Such states need only
have a small number of particles in the states that are entangled.

As summarised in Section II, the observation $E_{HZ}<1$ implies entanglement
between the two modes (and hence the two groups of atoms), which we
will refer to as $a$ and $b$ in keeping with the notation for the
associated boson operator symbols. \textcolor{black}{Our first result
is as follows:}

\textbf{\textcolor{black}{\emph{Result (1):}}}\textcolor{black}{{}
If $E_{HZ}$ is measured experimentally, one can} determine the maximum
value $s_{0}$ such that the following holds: 
\begin{equation}
\frac{E_{HZ}}{r}<\tilde{C}_{s_{0}}\label{eq:proc}
\end{equation}
where $r=\frac{|\langle\vec{S}\rangle|}{\langle\hat{N}\rangle/2}$.
Here we introduce the Bloch vector $\langle\vec{S}\rangle=(\langle\hat{S}_{x}\rangle,\langle\hat{S}_{y}\rangle,\langle\hat{S}_{z}\rangle)$.
Hence $|\langle\vec{S}\rangle|=\sqrt{\langle\hat{S}_{x}\rangle^{2}+\langle\hat{S}_{y}\rangle^{2}+\langle\hat{S}_{z}\rangle^{2}}$.
We restrict to regimes where $|\langle\vec{S}\rangle|$ is measured
to be non-zero. The conclusion from the measurements is that the two-mode
entanglement depth is at least $2s_{0}$.

\textcolor{black}{The statement of Result (1) can be clarified for
the different contexts of pure and mixed states. If the system were
a }\textcolor{black}{\emph{pure}}\textcolor{black}{{} state, then
the conclusion is that the system is in a pure bosonic two-mode entangled
state which has a mean particle number $\hat{N}$ of at least $2s_{0}$.
If the system is in a probabilistic }\textcolor{black}{\emph{mixture}}\textcolor{black}{{}
of pure entangled and non-entangled states, then the conclusion is
that the system exists, with a nonzero probability $P_{R}$, in a
pure two-mode bosonic entangled state $|\psi_{R}\rangle$ of at least
$2s_{0}$ particles. }

\textcolor{black}{The criteria we derive in this paper apply to all
two-mode systems, including photonic systems, for which a mixed state
analysis is important. Bose-Einstein condensates prepared experimentally
have a high degree of purity, but are nonetheless subject to interactions
with the environment that result in a loss of atoms from the condensate.
There are hence fluctuations of the atom number $N$ of the condensate.
A complete treatment therefore requires consideration of mixed states.
Analyses of Bose-Einstein condensates often assume pure states with
a fixed atom number $N$. This would imply $P_{R}\sim1$.}

\textbf{\textcolor{black}{\emph{Proof}}}\textcolor{black}{{} }\textbf{\textcolor{black}{\emph{of
Result (1):}}}\textcolor{black}{{} The system is described by a density
matrix} $\rho=\sum_{R}P_{R}|\psi_{R}\rangle\langle\psi_{R}|$ where
$|\psi_{R}\rangle$ is a pure state and $P_{R}$ are probabilities
($\sum_{R}P_{R}=1$, $P_{R}>0$). Each $|\psi_{R}\rangle$ either
satisfies a separable model or not. We can write the density operator
in the form $\rho=P_{sep}\rho_{sep}+P_{ent}\rho_{ent}$ where $P_{sep}$,
$P_{ent}$ are probabilities such that $P_{sep}+P_{ent}=1$. Here
$\rho_{sep}$ is a density operator for states described by the separable
model. The entangled part of the density operator that does not satisfy
the separable model is written 
\begin{equation}
\rho_{ent}=\sum_{R'}P_{R'}|\psi_{R'}\rangle\langle\psi_{R'}|\label{eq:4-1}
\end{equation}
where $\sum_{R'}P_{R'}=1$ and each $|\psi_{R'}\rangle$ is an entangled
pure two-mode state with $n_{R'}$ particles.\textcolor{black}{{}
The expression for $\rho$ that gives the decomposition into a separable
and nonseparable part is not required to be unique, as the following
proof holds for any such decomposition. Genuine lower bounds can thus
be established. }

\textcolor{black}{For a mixture the following is true \cite{hofmann2}}
\begin{eqnarray}
(\Delta\hat{S}_{x})^{2}+(\Delta\hat{S}_{y})^{2} & \geq & \sum_{R}P_{R}\{(\Delta_{R}\hat{S}_{x})^{2}+(\Delta_{R}\hat{S}_{y})^{2}\}\nonumber \\
\label{eq:mixproof}
\end{eqnarray}
where $(\Delta_{R}\hat{S}_{x})^{2}+(\Delta_{R}\hat{S}_{y})^{2}$ is
the sum of the variances for the pure state $|\psi_{R}\rangle$. Each
state $|\psi_{R}\rangle$ may be written as a linear combination of
spin eigenstates $|Sm\rangle$ of $\hat{S}^{2}$ and $\hat{S}_{z}$
(which form a basis). We note however that where $|\psi_{R}\rangle$
is a superposition of states with different $S$, the averages $\langle\hat{S}^{2}\rangle$
and $\langle\hat{S}_{x/y}\rangle$ are equal to those of the corresponding
mixtures (because states with different $S$ will be orthogonal) and
hence we do not treat this as a special case: It suffices to take
a fixed $s_{R}$ for each $|\psi_{R}\rangle$.

We next denote $s_{0}$ as the maximum value of the set $\{s_{R'}\neq0\}$
over the \emph{entangled} states. If all $s_{R'}=0$, then we take
$s_{0}=1/2$. Some states may have a zero spin $s_{R}=0$. However,
we need only consider the sum over states with $s_{R}\neq0$ and use
the definition $\tilde{C}_{s}=C_{s}/s$, to write: 
\begin{eqnarray}
(\Delta\hat{S}_{x})^{2}+(\Delta\hat{S}_{y})^{2} & \geq & \sum_{R}P_{R}\{\Delta_{R}(\hat{S}_{x})^{2}+(\Delta_{R}\hat{S}_{y})^{2}\}\nonumber \\
 & \geq & \sum_{R}P_{R}s_{R}\tilde{C}_{s_{R}}\label{eq:eqaray}
\end{eqnarray}
The first step remains valid with the restriction to $R$ such that
$s_{R}\neq0$. In the second step, and in all summations over $R$
written below, we take this restriction as implicit. Now we apply
the result $\frac{(\Delta\hat{S}_{x})^{2}+(\Delta\hat{S}_{y})^{2}}{|\langle\vec{S}\rangle|}\geq1$
that holds for the separable states, based on $|\langle\vec{S}\rangle|\leq\langle\hat{N}\rangle/2$
and that separable states satisfy $E_{HZ}\geq1$. We find 
\begin{eqnarray*}
(\Delta\hat{S}_{x})^{2}+(\Delta\hat{S}_{y})^{2} & \geq & P_{sep}\sum_{R''}P_{R''}s_{R''}\\
 &  & +P_{ent}\sum_{R'}P_{R'}s_{R'}\tilde{C}_{s_{R'}}
\end{eqnarray*}
The $C_{S}/S$ functions are monotonically decreasing with $S\neq0$.
This implies that $\sum_{R'}P_{R'}s_{R'}\tilde{C}_{s_{R'}}\geq\tilde{C}_{s_{0}}\sum_{R'}P_{R'}s_{R'}$.
Hence 
\begin{eqnarray*}
(\Delta\hat{S}_{x})^{2}+(\Delta\hat{S}_{y})^{2} & \geq & P_{sep}\sum_{R''}P_{R''}s_{R''}\\
 &  & +P_{ent}\tilde{C}_{s_{0}}\sum_{R'}P_{R'}s_{R'}
\end{eqnarray*}
Hence 
\begin{eqnarray}
(\Delta\hat{S}_{x})^{2}+(\Delta\hat{S}_{y})^{2} & \geq & \tilde{C}_{s_{0}}\sum_{R}P_{R}s_{R}\nonumber \\
 & \geq & \tilde{C}_{s_{0}}|\langle\vec{S}\rangle|\label{eq:array5}
\end{eqnarray}
In the last step we use $\sum_{R}P_{R}s_{R}\geq\sum_{R}P_{R}|\langle\hat{S}_{\theta,\phi}\rangle_{R}|$
where $\langle\hat{S}_{\theta,\phi}\rangle_{R}$ (for the state denoted
$R$) is the mean of an arbitrary spin component denoted by $\hat{S}_{\theta,\phi}$,
including the $\theta$ and $\phi$ that define the orientation of
the Bloch vector $\vec{S}_{R}=(\langle\hat{S}_{x}\rangle_{R},\langle\hat{S}_{y}\rangle_{R},\langle\hat{S}_{z}\rangle_{R})$.
This implies $\sum_{R}P_{R}s_{R}\geq\sum_{R}P_{R}|\vec{S}{}_{R}|\geq|\sum_{R}P_{R}\vec{S}{}_{R}|$
and hence $\sum_{R}P_{R}s_{R}\geq|\langle\vec{S}\rangle|$ where $\langle\vec{S}\rangle=\left(\langle\hat{S}_{x}\rangle,\langle\hat{S}_{y}\rangle,\langle\hat{S}_{z}\rangle\right)=\sum_{R}P_{R}\vec{S}_{R}$
is the Bloch vector defined for the state $\rho$. Using the definition
of $r$ given by (\ref{eq:proc}), we obtain 
\begin{equation}
E_{HZ}\geq r\tilde{C}_{j_{0}}\label{eq:ent-number}
\end{equation}
Thus, if we measure $E_{HZ}<r\tilde{C}_{s_{0}}$, we deduce that one
of the pure entangled states $|\psi_{R'}\rangle$ must possess a spin
greater than $s_{0}$. The number of particles $n_{R'}$ in this state
$|\psi_{R'}\rangle$ is more than $2s_{0}$. This completes the proof.

\subsection{Depth of two-mode EPR steering}

We note from Figure 1 that in fact $C_{S}<0.5$ for $S>1/2$. It is
thus possible to extend Result (1) to include EPR steerable states.
We define the \emph{two-mode EPR steering depth} as the number of
boson particles that comprise a pure two-mode EPR steerable state.
Next we give the main result of the paper.

\textbf{\emph{Result 2:}} If the experiment reveals $E_{HZ}<0.5$
, so that we can identify a value $s_{0}$ such that 
\begin{equation}
E_{HZ}<r\tilde{C}_{s_{0}}<0.5\label{eq:mainresult}
\end{equation}
where $r=\frac{|\langle\vec{S}\rangle|}{\langle\hat{N}\rangle/2}$,
then we deduce a two-mode EPR steering depth of at least $2s_{0}$.
\textcolor{black}{If the system were a pure state, the statement means
that there is a minimum of $2s_{0}$ particles in the pure two-mode
EPR steerable state. If the system is in a mixture $\rho$, then the
statement means that (necessarily) there is a nonzero probability
$P_{R}$ for the system being in a pure EPR steerable state $|\psi_{R}\rangle$
with at least $2s_{0}$ particles.}\textbf{\textcolor{black}{{} }}

\textbf{Proof of Result (2):} We extend the previous proof. In any
decomposition of the density operator $\rho=\sum_{R}P_{R}|\psi_{R}\rangle\langle\psi_{R}|$,
each $|\psi_{R}\rangle$ either satisfies LHS models (\ref{eq:lhs})
and (\ref{eq:lhs-1}) (and is therefore non-steerable), or not. We
can write the density operator in the form $\rho=P_{lhs}\rho_{lhs}+P_{st}\rho_{st}$
where $P_{lhs}$, $P_{st}$ are probabilities such that $P_{lhs}+P_{st}=1$.
Here $\rho_{lhs}$ is a density operator for states described by the
LHS models, which includes all separable states. The steerable part
of the density operator that does not satisfy both LHS models (\ref{eq:lhs})
and (\ref{eq:lhs-1}) is written 
\begin{equation}
\rho_{st}=\sum_{R'}P_{R'}|\psi_{R'}\rangle\langle\psi_{R'}|\label{eq:4}
\end{equation}
where $\sum_{R'}P_{R'}=1$. Here each $|\psi_{R'}\rangle$ is an EPR
steerable pure two-mode state. Following the proof of Section III.A,
we denote $s_{0}$ as the maximum value of the set $\{s_{R}\neq0\}$
over the \emph{steerable} states. If all $s_{R'}=0$, we take $s_{0}=1/2$.
Some states may have a zero spin $s_{R}=0$. However, we consider
the sum over states with $s_{R}\neq0$ and use the definition $\tilde{C}_{s}=C_{s}/s$,
to write, following the lines (\ref{eq:eqaray}), $(\Delta\hat{S}_{x})^{2}+(\Delta\hat{S}_{y})^{2}\geq\sum_{R}P_{R}\{(\Delta_{R}\hat{S}_{x})^{2}+(\Delta_{R}\hat{S}_{y})^{2}\}$.
Hence 
\begin{equation}
(\Delta\hat{S}_{x})^{2}+(\Delta\hat{S}_{y})^{2}\geq\sum_{R}P_{R}s_{R}\tilde{C}_{s_{R}}\label{eq:314}
\end{equation}
For non-steerable states (which imply that both LHS models (\ref{eq:lhs})
and (\ref{eq:lhs-1}) hold), we know from Section II that $E_{HZ}\geq0.5$.
Thus, $\frac{\left(\Delta\hat{S}_{x}\right)^{2}+\left(\Delta\hat{S}_{y}\right)^{2}}{|\langle\vec{S}\rangle|}\geq0.5$
must hold for the separable and nonsteerable states. We find 
\begin{eqnarray*}
(\Delta\hat{S}_{x})^{2}+(\Delta\hat{S}_{y})^{2} & \geq & 0.5P_{lhs}\sum_{R'}P_{R''}s_{R''}\\
 &  & +P_{st}\sum_{R'}P_{R'}s_{R'}\tilde{C}_{s_{R'}}
\end{eqnarray*}
Since $\tilde{C}_{s_{R'}}\geq\tilde{C}_{s_{0}}$, this becomes 
\begin{eqnarray*}
(\Delta\hat{S}_{x})^{2}+(\Delta\hat{S}_{y})^{2} & \geq & 0.5P_{lhs}\sum_{R''}P_{R''}s_{R''}\\
 &  & +P_{st}\tilde{C}_{j_{0}}\sum_{R'}P_{R'}s_{R'}
\end{eqnarray*}
Since $\tilde{C}_{s_{0}}\leq0.5$

\begin{eqnarray}
(\Delta\hat{S}_{x})^{2}+(\Delta\hat{S}_{y})^{2} & \geq & \tilde{C}_{s_{0}}\sum_{R}P_{R}s_{R}\nonumber \\
 & \geq & \tilde{C}_{s_{0}}|\langle\vec{S}\rangle|\label{eq:f}
\end{eqnarray}
This implies $E_{HZ}\geq r\tilde{C}_{s_{0}}$. Thus, if we measure
$E_{HZ}<r\tilde{C}_{s_{0}}$, we deduce an EPR steerable state with
spin greater than $s_{0}$, and thus an EPR steerable state $|\psi_{R'}\rangle$
with a total number $n_{R'}$ of bosons of more than $2s_{0}$. The
conclusion is that there is a minimum of $n_{st}=2s_{0}$ bosons involved
in the two-mode EPR steerable state.\textcolor{black}{{} This completes
the proof.}

\subsection{Depth of two-mode EPR steering based on PQS}

\textcolor{black}{It is possible to obtain more sensitive criteria
for the depth of two-mode steering by considering the lower bounds,
derived recently by Vitagliano et al. \cite{toth-planar-ent}, of
$(\Delta\hat{S}_{x})^{2}+(\Delta\hat{S}_{y})^{2}$, for a given $S$
}\textcolor{black}{\emph{and}}\textcolor{black}{{} $\langle S_{||}\rangle=\sqrt{\langle\hat{S}_{x}\rangle^{2}+\langle\hat{S}_{y}\rangle^{2}}$.
These authors applied the bounds to deduce large numbers of atoms
entangled in an atomic thermal ensemble. In particular, Vitagliano
et al. derive convex functions $F_{S}^{(1/2)}$ such that for a fixed
$S$ 
\begin{equation}
\frac{(\Delta\hat{S}_{x})^{2}+(\Delta\hat{S}_{y})^{2}}{\langle\hat{N}\rangle/2}\geq F_{S}^{(1/2)}\Bigl(\frac{\langle S_{||}\rangle}{\langle\hat{N}\rangle/2}\Bigr)\label{eq:planefuns-2}
\end{equation}
Here we use the superscript $(1/2)$ to indicate we restrict to spin
$1/2$ particles i.e. each particle has two levels (modes) available
to it. We prove the following. }

\textbf{\textcolor{black}{\emph{Result 3:}}}\textcolor{black}{{}
If the measurement of $E_{HZ}$ and $r_{||}=\frac{|\langle S_{||}\rangle|}{\langle N\rangle/2}$
yield values such that $E_{HZ}<F_{s_{0}}(\frac{|\langle S_{||}\rangle|}{\langle\hat{N}\rangle/2})$
where $F_{S}(x)\leq0.5$ for all $0\leq x\leq1$, and if the functions
$F_{S}(x)$ are monotonically decreasing with $S$ for every fixed
$0\leq x\leq1$, then the }\textcolor{black}{\emph{EPR steering depth}}\textcolor{black}{{}
is (at least) $2s_{0}$. The proof is a straightforward extension
of the proofs given in III.B and is given in the Appendix A. }

\textcolor{black}{One suitable lower bound are the functions considered
by Vitagliano et al of $F_{S}^{(1/2)}(x)=x\zeta_{S}^{2}$, where $\zeta_{S}^{2}$
is the minimum planar spin squeezing value $\xi_{||}^{2}=\frac{\left(\Delta S_{y}\right)^{2}+\left(\Delta S_{z}\right)^{2}}{\left|\left\langle S_{||}\right\rangle \right|}$
over all single particle states $|\psi_{k}\rangle$ of spin $S=k/2$
. Substituting into (\ref{eq:planefuns-2}) leads to the condition
\begin{equation}
\frac{E_{HZ}}{r_{||}}<\zeta_{s_{0}}^{2}\label{eq:proc-1}
\end{equation}
for a two-mode $s_{0}$-particle EPR steering depth. Examination of
the functions $\zeta_{S}^{2}$ evaluated in Ref. \cite{toth-planar-ent}
reveal similarity with the $\tilde{C}_{S}$ functions (and the Result
2 of Section III.B) associated with the fact that the planar squeezed
states minimising $C_{S}$ have Bloch vector orientated along $\vec{S}_{x}$
\cite{cj-2,hesteer-1}. The values are seen to be monotonically decreasing
with $S$ and satisfy $\zeta_{S}^{2}\leq0.5$, implying the conditions
necessary for the Result 3 ($r_{||}\leq1$). }

\section{two-mode BEC interferometer}

\subsection{Interaction}

To illustrate the usefulness of the criteria, we consider a simple
model for a two-mode interferometer. For convenience, we symbolise
the systems $A$ and $B$ by the notation for the associated boson
operators: $a$ and $b$. The two modes become entangled when an initial
state consisting of a number state $|N\rangle_{a}$ in mode $a$ and
a vacuum state $|0\rangle_{b}$ in mode $b$ are coupled by a $50/50$
beam splitter (Figure 2) \cite{hesteer-1,noonmesopaper}. The state
generated after the interaction is 
\begin{equation}
|\psi\rangle=\sum_{r}c_{r}|N-r\rangle_{a}|r\rangle_{b}\label{eq:bsstate-1}
\end{equation}
where $c_{r}=\sqrt{N!}/\sqrt{2^{N}r!(N-r)!}$ \cite{noonmesopaper}.\textcolor{red}{{}
}The state $|\psi\rangle$ is entangled. This can be certified experimentally
using the HZ entanglement criterion, $E_{HZ}<1$ \cite{hesteer-1}.
It was shown in reference \cite{hesteer-1} that for $|\psi\rangle$,
$E_{HZ}\rightarrow0.5$ as $N\rightarrow\infty$. This result is plotted
in Figure 3. \textcolor{black}{Here $\langle\hat{a}^{\dagger}\hat{a}\rangle=\langle\hat{b}^{\dagger}\hat{b}\rangle$
and hence the steering criterion given by (\ref{eq:ehzsteer}) is
$E_{HZ}<0.5$, which is not achieved} for the simple beam splitter
interaction.

\textcolor{black}{The transformations illustrated in Figure 2 also
apply to a two-mode BEC atom interferometer. For the details of such
interferometers, the reader is referred to Refs. \cite{yun-li,yunli-2,Egorov-1,Philipp2010-1,bjdtwo-mode,bogdanepl}.
Here, $N$ atoms are prepared as a single component BEC in the hyperfine
atomic level denoted $|1\rangle$. A second atomic hyperfine level
is denoted $|2\rangle$. The beam splitter interaction symbolised
$BS1$ in the Figure is achieved by a Rabi rotation. This involves
application of a $\pi/2$ microwave pulse to the atomic ensemble,
to prepare the atoms in a two-component BEC which is in a superposition
of the two atomic levels. In a two-mode model, the components of the
BEC in levels $|1\rangle$ and $|2\rangle$ are associated with stationary
mode functions which we identify respectively as modes $a$ and $b$.
The atoms are no longer distinguishable particles but are $N$ bosons
of a condensate mode. Comparisons with real interferometers show good
agreement with experiment in suitable parameter limits \cite{yunli-2}.
The two-mode model is relevant only at low temperatures below the
critical value where the thermal fraction is negligible. After the
interaction denoted $BS1$, the state given by $|\psi\rangle$ can
be represented on a Bloch sphere as a spin coherent state. Here, the
Bloch vector is aligned along the $x$ direction and has a magnitude
$N/2$, and the variances in the $yz$ plane are equal ($(\Delta\hat{S}_{y})^{2}=(\Delta\hat{S}_{z})^{2}$). }

\textcolor{black}{The variances required for the $E_{HZ}$ criteria
can be measured in terms of number differences at the output of the
BEC interferometer. The interferometer has a second beam splitter
$BS2$ with the two single mode inputs $a$ and $b$ (Figure 2). Introducing
a relative phase shift $\phi$, the boson destruction operators of
the output modes of the second beam splitter are $\hat{c}=(\hat{a}-\hat{b}\exp^{i\phi})/\sqrt{2}$,
$\hat{d}=(\hat{a}+\hat{b}\exp^{i\phi})/\sqrt{2}$. In the BEC interferometer,
the second beam splitter is realised as a second microwave pulse \cite{Philipp2010-1,Egorov-1}.
The output n}umber difference operator is given as $\hat{M}=\hat{d}^{\dagger}\hat{d}-\hat{c}^{\dagger}\hat{c}=2\hat{S}_{x}\cos\phi+2\hat{S}_{y}\sin\phi$.
Selecting $\phi=0$ or $\phi=\pi/2$ enables measurements of $\hat{S}_{y}$
or $\hat{S}_{x}$. The $\langle\hat{S}_{z}\rangle$ and $(\Delta\hat{S}_{z})^{2}$
can be measured directly without the second beam splitter, or by passing
the outputs $c$ and $d$ through a second transformation with $\phi=0$.
\begin{figure}
\includegraphics[width=0.8\columnwidth]{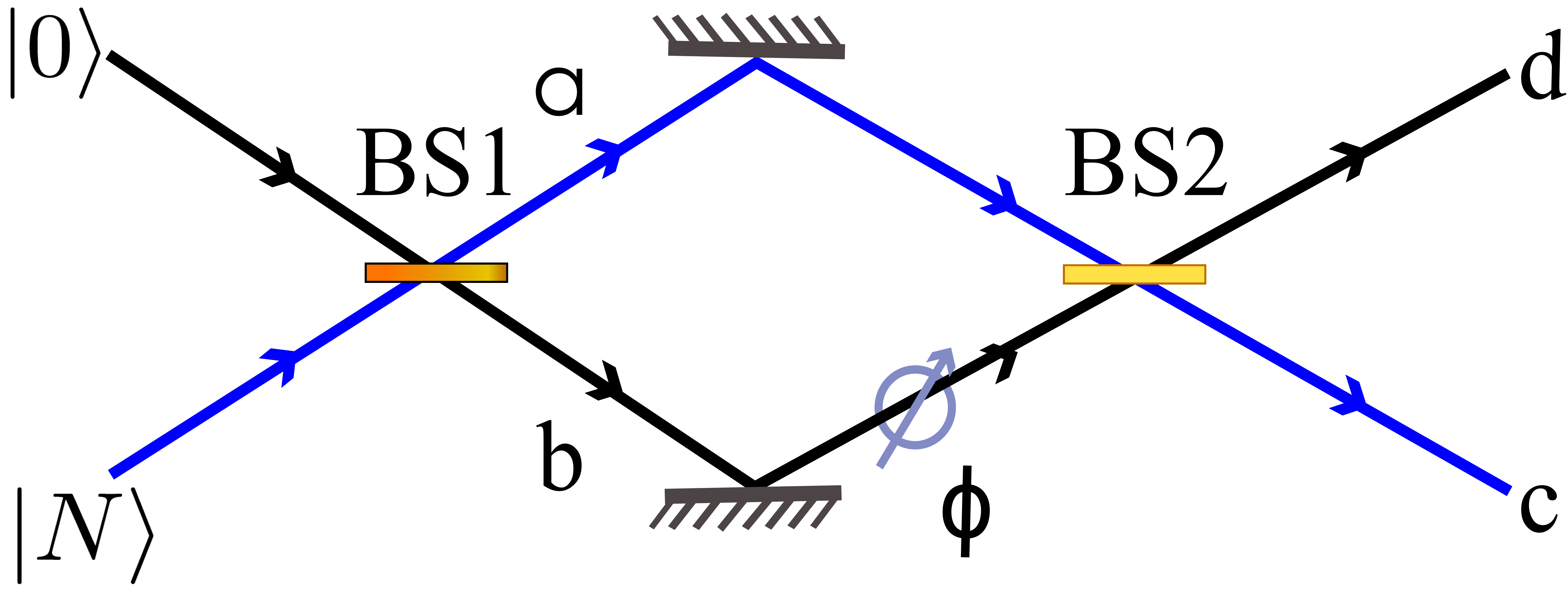}

\caption{\textcolor{black}{Entangled modes $a$ and $b$ are created when a
number state $|N\rangle$ is incident on the beam splitter $BS1$.
The entanglement can be detected when the modes $a$ and $b$ interfere
across a beam splitter $BS2$ with a phase shift $\varphi$. The two-mode
number difference $\hat{M}$ measured at the outputs depends on $\phi$,
which enables measurement of the variances $\hat{S}_{x}$, $\hat{S}_{z}$
and $\hat{S}_{x}$. A nonlinear medium may be present after the first
beam splitter $BS1$, as modelled by the nonlinear Hamiltonian $H_{NL}$.
The description of how the beam splitters are implemented for a BEC
interferometer is given in the text.}}
\end{figure}

\begin{figure}
\includegraphics[width=0.6\columnwidth]{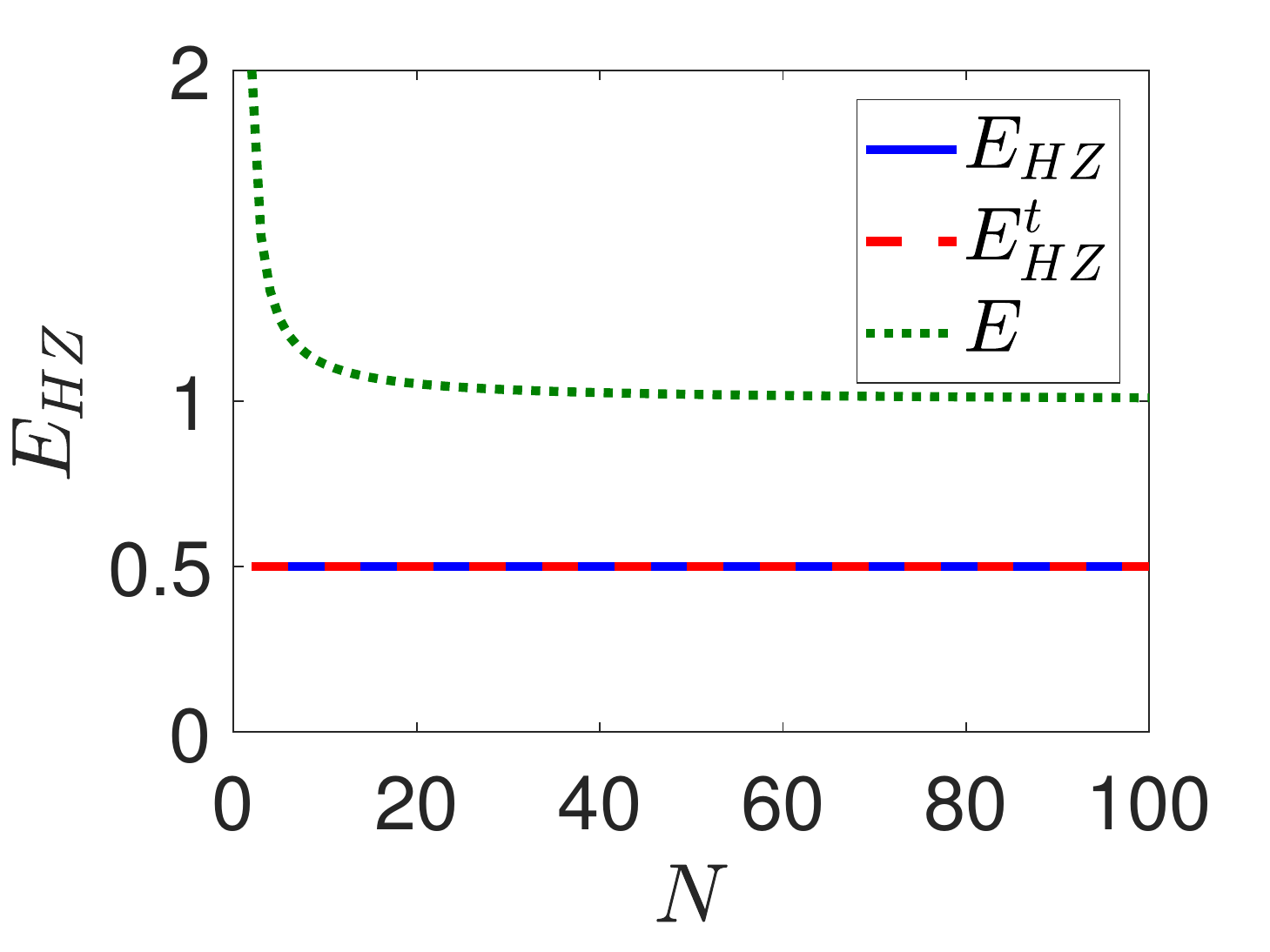}

\caption{\textcolor{black}{Entanglement between the two modes $a$ and $b$
of the interferometer of Figure 2 in the absence of nonlinearity ($\chi=0$)
is detectable by measurement of the Hillary-Zubairy parameter $E_{HZ}$.
Also plotted is the value $E_{HZ}^{t}$ for the rotated spin vectors
defined in the text, and the ratio $E=|\langle\hat{a}^{\dagger}\hat{b}\rangle|^{2}/\langle\hat{a}^{\dagger}\hat{a}\hat{b}^{\dagger}\hat{b}\rangle$.
There is entanglement between the two modes $a$ and $b$ if $E_{HZ}<1$
or $E>1$. $E_{HZ}^{t}<1$ certifies entanglement between the rotated
modes defined by Eq. (\ref{eq:ehzrotz}).}}
\end{figure}

In order to model the nonlinearity of the atomic medium in a BEC interferometer,
we consider that subsequent to the initial Rabi rotation (modelled
by $BS1$) the system evolves for a time $t$ according to a nonlinear
Hamiltonian\textcolor{black}{{} 
\begin{equation}
H_{NL}=\chi(\hat{a}^{\dagger2}\hat{a}^{2}+\hat{b}^{\dagger2}\hat{b}^{2}+2K\hat{a}^{\dagger}\hat{a}\hat{b}^{\dagger}\hat{b}+\hat{a}^{\dagger}\hat{a}+\hat{b}^{\dagger}\hat{b})\label{eq:fullnlk}
\end{equation}
Here $K$ is a constant is adjusted to model different atomic interferometers.
For $K=-1$ the Hamiltonian reduces to} $H_{NL}=\chi(\hat{a}^{\dagger}\hat{a}-\hat{b}^{\dagger}\hat{b})^{2}$\textcolor{black}{{}
as studied in Refs \cite{yun-li}. This Hamiltonian is a generalised
form of the well-known Josephson Hamiltonian (see ref \cite{bjdtwo-mode}
for a discussion) based on assuming both the mode functions and their
occupancy remain fixed. We may also allow $K=0$ to model the interaction
$H_{NL}=\chi(\hat{a}^{\dagger}\hat{a})^{2}+(\hat{b}^{\dagger}\hat{b})^{2}$.
This interaction is an approximation to the multi-mode BEC interferometer
discussed in Ref. \cite{bogdanepl}. After a time $t$ the state $|\psi\rangle$
evolves to}

\textcolor{black}{{} 
\begin{eqnarray}
|\psi(t)\rangle & = & \sum_{r-0}^{\infty}c_{r}e^{-i\Omega(r)t/\hbar}|N-r\rangle_{a}|r\rangle_{b}\label{eq:nlsoln}
\end{eqnarray}
where $\Omega(r)=\chi\left[(N-r)^{2}+r^{2}+2Kr(N-r)\right]$ is the
term due to nonlinearity. After a time $t$, the second Rabi rotation
explained above allows measurement of the spins $\hat{S}_{x}$, $\hat{S}_{y}$
and $\hat{S}_{z}$ and their variances. In the atom interferometer,
the number difference is measured by atom imaging techniques.}

\subsection{Evolution of $\hat{S}_{x}$, $\hat{S}_{y}$ and $\hat{S}_{z}$}

\textcolor{black}{In Figure 4, we plot $E_{HZ}$ and the spin variances
for $N=100$, versus $t$ for both $K=-1$ and $K=0$. The solutions
show that the Bloch vector is orientated along $\hat{S}_{x}$ ($\langle\hat{S}_{y}\rangle=\langle\hat{S}_{z}\rangle=0$).
Initially, $\langle\hat{S}_{x}\rangle\sim N/2$. We plot the evolution
of $\langle\hat{S}_{x}\rangle$ noting the drop in value with time.
For small times, the solutions show a noise reduction in $\hat{S}_{x}$
with $\Delta\hat{S}_{x}\rightarrow0$ at $t\rightarrow0$. Spin squeezing
is defined when $(\Delta\hat{S}_{z})^{2}<|\langle\hat{S}_{x}\rangle|/2$
or $(\Delta\hat{S}_{y})^{2}<|\langle\hat{S}_{x}\rangle|/2$. The plots
show there is no spin squeezing in $\hat{S}_{z}$ or $\hat{S}_{y}$.
The moment $\langle\hat{S}_{z}^{2}\rangle$ is independent of time
and also of $\Omega$, for both $K=0$ and $K=-1$. In fact, we find
that $(\Delta\hat{S}_{z})^{2}=N/4$. }The Figures show that the variance
in $\hat{S}_{y}$ can exceed the level of $N/4$.\textcolor{black}{{} }

\begin{figure}
\includegraphics[width=0.5\columnwidth]{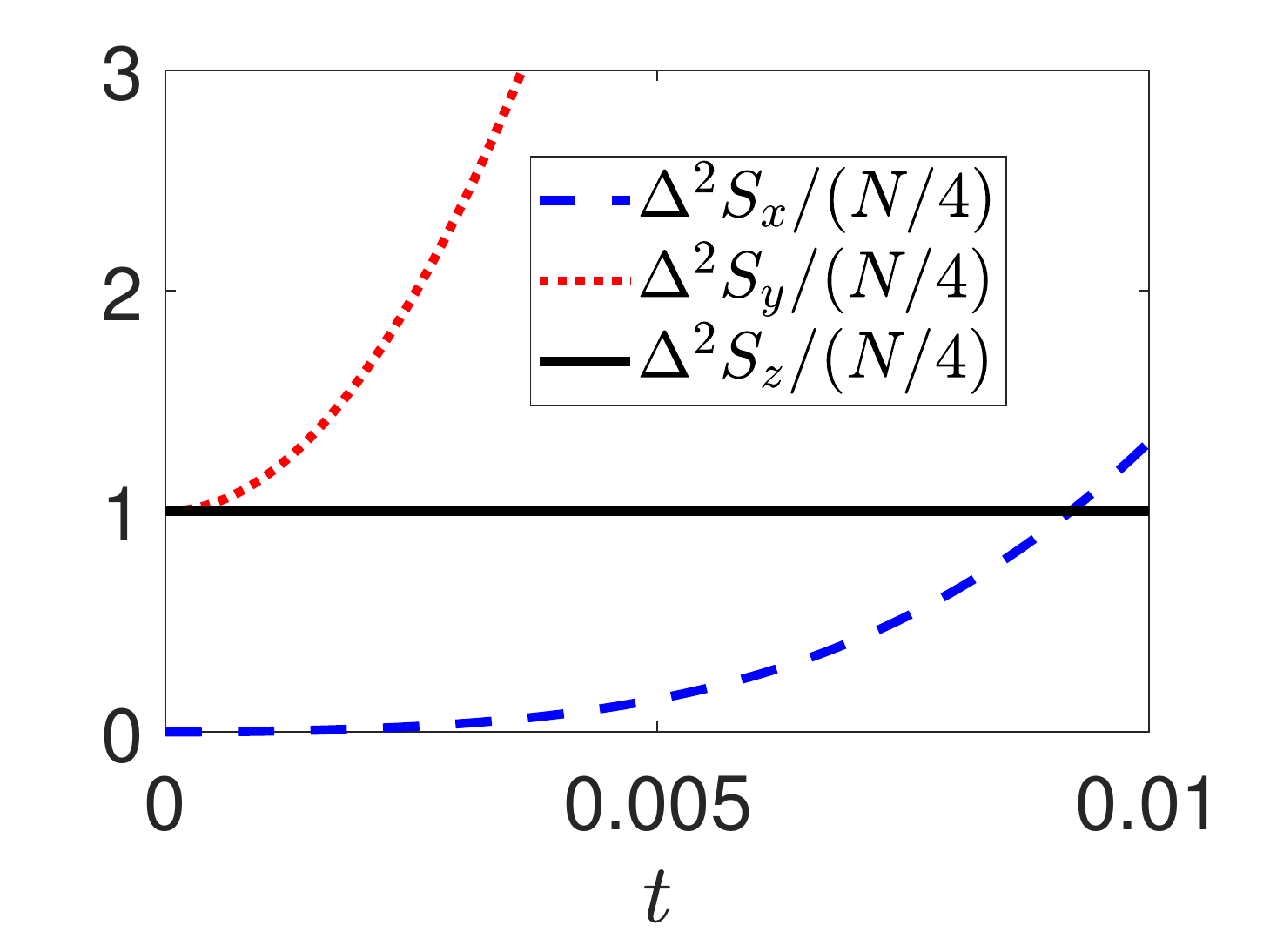}\includegraphics[width=0.5\columnwidth]{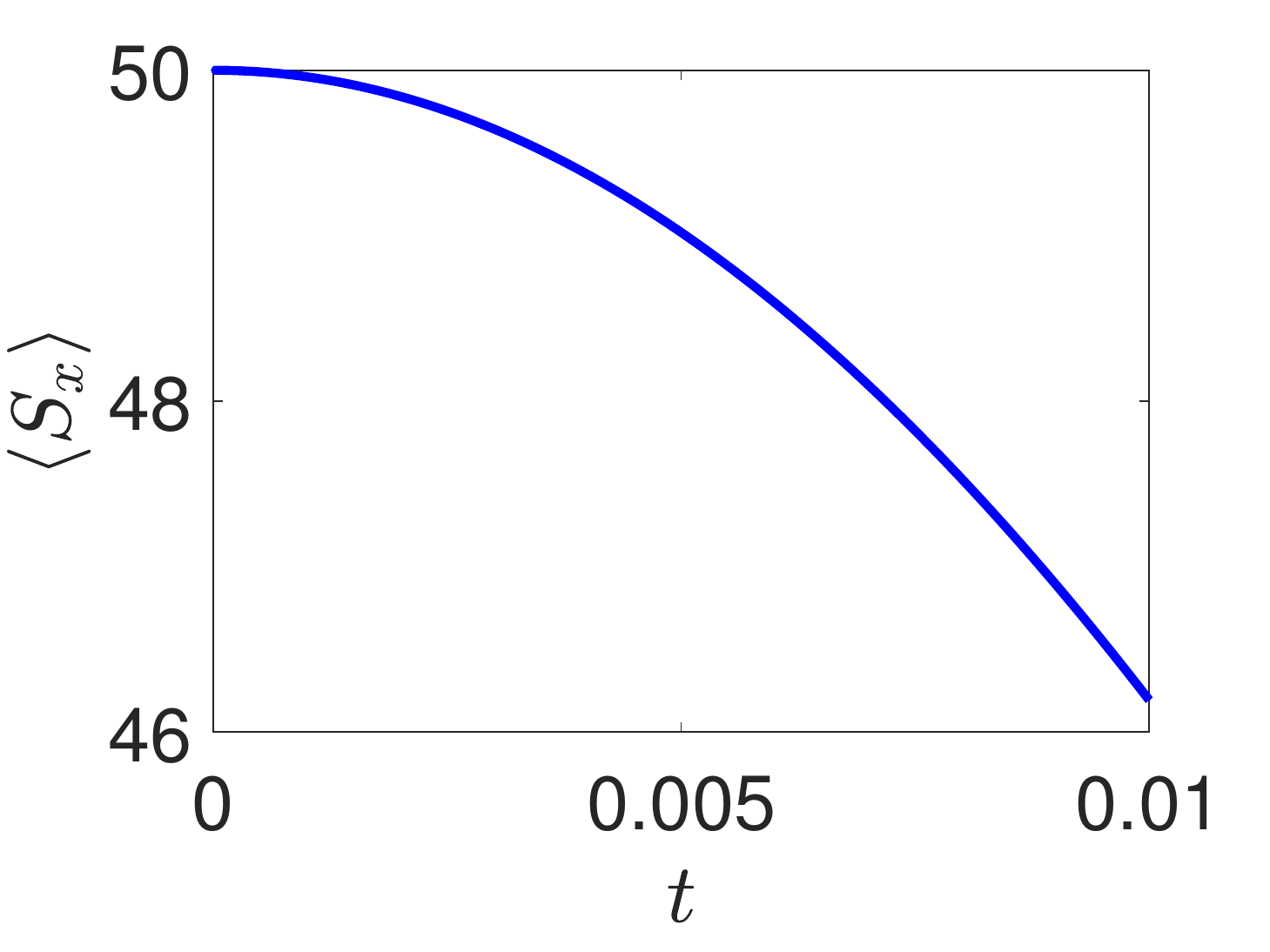}

\includegraphics[width=0.5\columnwidth]{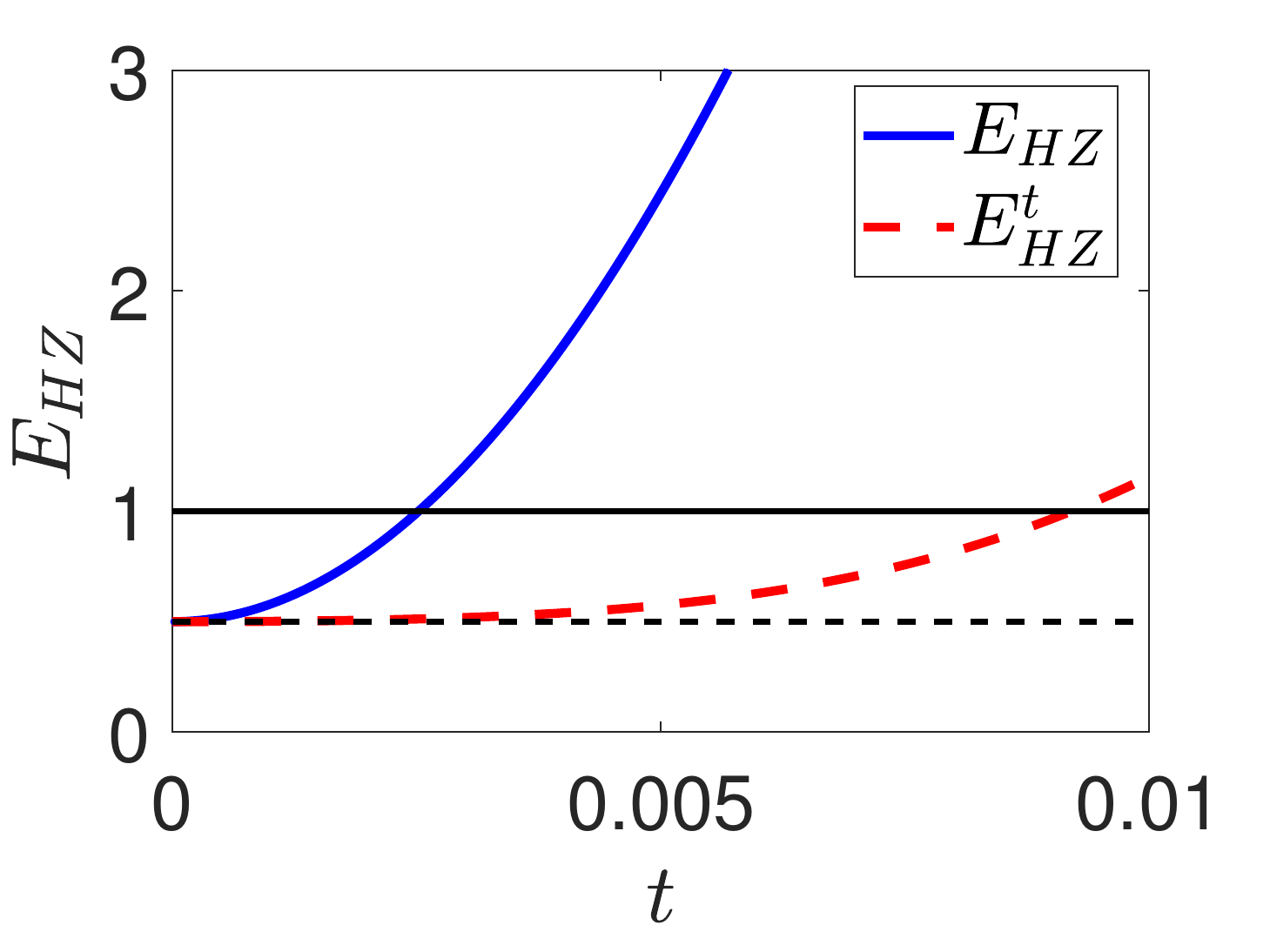}

\caption{\textcolor{black}{The graphs show the spin variances and mean spin
values for the fields created in the nonlinear interferometer of Figure
2 after a time $t$ with nonlinearity present ($\chi\protect\neq0$).
Here time $t$ is in units of $1/\chi$. $N=100$ and $K=-1$. The
top left graph gives the variances of $\hat{S}_{x}$, $\hat{S}_{y}$
and $\hat{S}_{z}$. The top right graph shows $\langle\hat{S}_{x}\rangle$.
The lower graph gives $E_{HZ}$ and $E_{HZ}^{t}$.{} }\textcolor{red}{{} }\textcolor{black}{Entanglement
is signified if $E_{HZ}<1$ or $E_{HZ}^{t}<1$. The plots for $K=0$
are similar but with time in units of $2\chi$.}}
\end{figure}

\begin{figure}
\includegraphics[width=0.5\columnwidth]{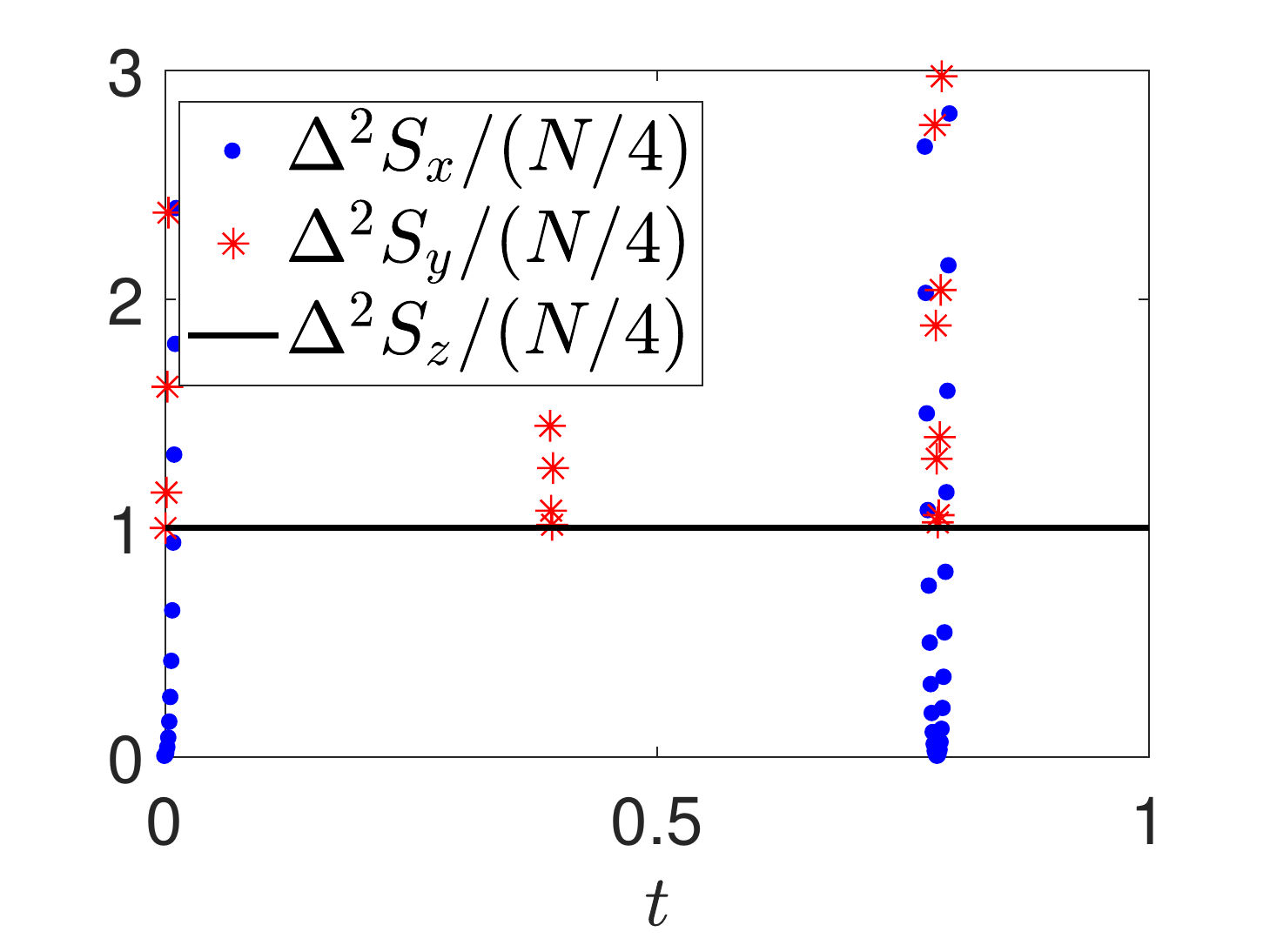}\includegraphics[width=0.5\columnwidth]{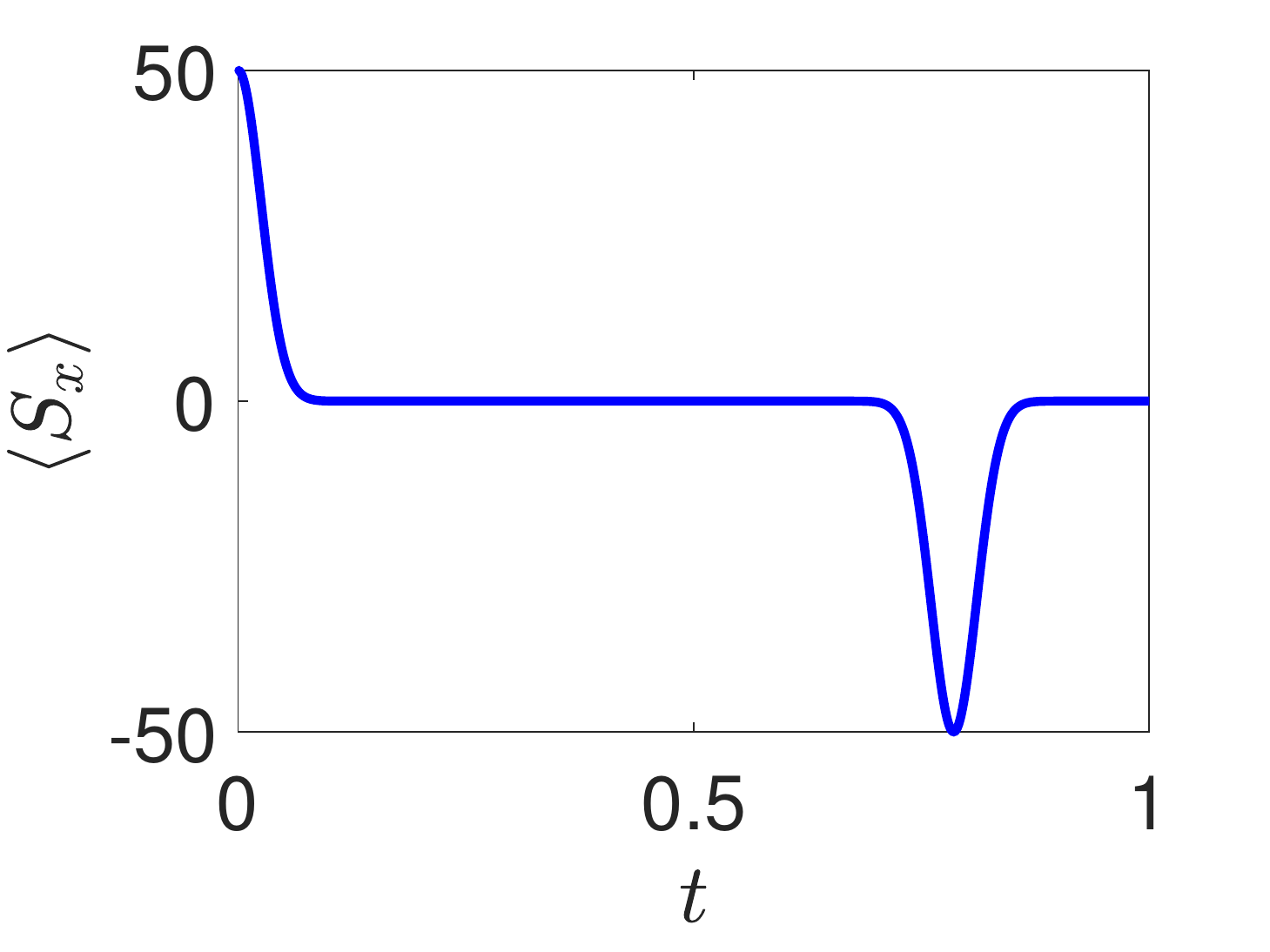}

\caption{\textcolor{black}{The graphs are as for Figure 4 showing the periodic
behaviour that is evident over longer time scales. }}
\end{figure}

\textcolor{black}{Figure 4 also plots the Hillery-Zubairy parameter
$E_{HZ}$. The $E_{HZ}$ entanglement value reduces below $1$, due
to the smallness of the variance $(\Delta\hat{S}_{x})^{2}$, which
is due to the precise number of atoms in the input state $|N\rangle$.
This enables a certification of entanglement between modes $a$ and
$b$. For longer times, the variances $(\Delta\hat{S}_{x})^{2}$ and
$(\Delta\hat{S}_{y})^{2}$ increase sufficiently to destroy the entanglement
signature. We also define a rotated $E_{HZ}$ parameter in terms of
a different planar spin squeezing orientation, as 
\begin{equation}
E_{HZ}^{t}=\frac{(\Delta\hat{S}_{x})^{2}+(\Delta\hat{S}_{z})^{2}}{\hat{N}/2}\label{eq:ehzrotz}
\end{equation}
}Details of the mode transformation assoicated with $E_{HZ}^{t}$
are given in Ref. \cite{hesteer-1}. We notice that due to the noise
reduction in $\hat{S}_{x}$, both \textcolor{black}{$E_{HZ}<1$ and
$E_{HZ}^{t}<1$ for smaller times. The signature for $E_{HZ}^{t}$
lasts longer than that of $E_{HZ}$, due to the stability of the variance
$(\Delta S_{z})^{2}$. The solutions for the nonlinear Hamiltonian
are periodic as evident from Figure 5, and for longer times there
is a return of the entanglement signature $E_{HZ}<1$ and $E_{HZ}^{t}<1$
coinciding with the vector $|\langle\hat{S}_{x}\rangle|\rightarrow N/2$.}

\subsection{\textcolor{black}{Spin squeezing of a spin vector in the $yz$ plane}}

\textcolor{black}{To optimise the detection of EPR-steering using
the Hillery-Zubairy entanglement criterion, we seek the optimal spin
squeezing for some $\hat{S}_{\theta}$ in the $yz$ plane. In that
case, where the variance $\Delta\hat{S}_{x}\rightarrow0$, the observation
of $(\Delta\hat{S}_{\theta})^{2}<N/4$ for $\theta$ in the $yz$
plane would imply an EPR-steering between two appropriately rotated
modes. In fact, such spin squeezing has been predicted for the two-mode
nonlinear Hamiltonian $H_{NL}$ by Li et al \cite{yun-li} and has
been observed experimentally \cite{Philipp2010-1}. With this motivation,
we define a spin vector in the $yz$ plane. Thus 
\begin{equation}
{\color{black}{\color{black}\hat{S}_{\theta}=\hat{S}_{y}\cos\theta+\hat{S}_{z}}\sin\theta}\label{eq:spintheta}
\end{equation}
Spin squeezing in $\hat{S}_{\theta}$ is observed when \cite{ueda-1}
\begin{equation}
(\Delta\hat{S}_{\theta})^{2}<|\langle\hat{S}_{x}\rangle|/2\label{eq:spinsq}
\end{equation}
We define the spin squeezing ratio 
\begin{equation}
\xi_{\theta}^{2}=\frac{(\Delta\hat{S}_{\theta})^{2}}{|\langle\hat{S}_{x}\rangle|/2}\label{eq:spinsqparameterwine}
\end{equation}
and note that where the Bloch vector is along the $x$ axis and $\langle\hat{S}_{x}\rangle\sim N/2$,
this is the definition $\bar{\xi}_{\theta}^{2}=\frac{\left\langle \hat{N}\right\rangle \left(\Delta\hat{S}_{\theta}\right)^{2}}{\left\langle \hat{S}_{x}\right\rangle ^{2}}$
used in Refs. \cite{ueda-1}. More generally, where $\langle\hat{S}_{x}\rangle<N/2$,
we see that $\bar{\xi}_{\theta}^{2}=\frac{\left\langle \hat{N}\right\rangle \xi_{\theta}^{2}}{2|\left\langle \hat{S}_{x}\right\rangle |}>\xi_{\theta}^{2}$
and spin squeezing as defined by $\xi_{\theta}^{2}<1$ does not imply
$\bar{\xi}_{\theta}^{2}<1$, though the converse is true.}\textcolor{red}{{}
}\textcolor{black}{Spin squeezing in $\hat{S}_{\theta}$ is observed
when $\xi_{\theta}^{2}<1$. Where $\langle\hat{S}_{x}\rangle=N/2$,
there is spin squeezing when $(\Delta\hat{S}_{\theta})^{2}<N/4$.
Although not evident in the plots for the $\hat{S}_{z}$ and $\hat{S}_{y}$
of Figures 4 and 5, it is know that spin squeezing is created for
optimal $\theta$ by the nonlinear dynamical evolution given by $H_{NL}$.
Spin squeezing is predicted by the simple model given by $H_{NL}$,
as has been shown in Refs \cite{yun-li,yunli-2}. }

\textcolor{black}{We summarise the calculation of Li et al. \cite{yun-li,yunli-2}.
We evaluate $(\Delta\hat{S}_{\theta})^{2}=\langle\hat{S}_{\theta}^{2}\rangle-\langle\hat{S}_{\theta}\rangle^{2}$.
Here the $\langle\hat{S}_{\theta}\rangle=0$ because $\langle\hat{S}_{y}\rangle=\langle\hat{S}_{z}\rangle=0$.
Therefore:}

\textcolor{black}{{} 
\begin{eqnarray}
\langle\hat{S}_{\theta}^{2}\rangle & = & \langle\cos\theta\hat{S}_{y}+\sin\theta\hat{S}_{z}\rangle\nonumber \\
 & = & \frac{1}{2}(\langle\hat{S}_{y}^{2}\rangle+\langle\hat{S}_{z}^{2}\rangle)-C\frac{\cos2\theta}{2}\nonumber \\
 &  & -\frac{i}{4}F\sin2\theta\label{eq:arrt}
\end{eqnarray}
}where we define 
\begin{eqnarray}
F & = & \langle\hat{a}^{\dagger2}\hat{a}\hat{b}-\hat{a}\hat{b}^{\dagger}-\hat{a}^{\dagger}\hat{a}^{2}\hat{b}^{\dagger}+\hat{b}^{\dagger}\hat{b}(\hat{a}\hat{b}^{\dagger}-\hat{a}^{\dagger}\hat{b})\rangle\nonumber \\
C & = & \langle\hat{S}_{z}^{2}\rangle-\langle\hat{S}_{y}^{2}\rangle\label{eq:arrty}
\end{eqnarray}
We wish to find the angle $\theta$ that produces the minimum value
of $\langle\hat{S}_{\theta}^{2}\rangle$. We \textcolor{black}{see
that}\textcolor{blue}{{} }

\begin{eqnarray}
\frac{\partial\langle\hat{S}_{\theta}^{2}\rangle}{\partial\theta} & = & C\sin2\theta-\frac{iF}{2}\cos2\theta=0\label{eq:min}
\end{eqnarray}
which implies the stationary condition $\tan2\theta=\frac{i}{2}\frac{F}{C}$.
Therefore the stationary values are at 
\begin{eqnarray}
\sin2\theta & = & \frac{\pm iF}{\sqrt{4C^{2}+\vert F\vert^{2}}}\nonumber \\
\cos2\theta & = & \frac{\pm2C}{\sqrt{4C^{2}+\vert F\vert^{2}}}\label{eq:soln}
\end{eqnarray}
On substituting into $\langle\hat{S}_{\theta}\rangle$ we find on
taking the minimum stationary value 
\begin{eqnarray}
\langle\hat{S}_{\theta}^{2}\rangle_{min} & = & \frac{1}{2}(\langle\hat{S}_{y}^{2}\rangle+\langle\hat{S}_{z}^{2}\rangle)-\frac{4C^{2}+\vert F\vert^{2}}{4\sqrt{4C^{2}+\vert F\vert^{2}}}\nonumber \\
\label{eq:mins}
\end{eqnarray}
\textcolor{black}{Following Li et al., we find significant spin squeezing
is possible for an optimal $\theta$ and time $t$. The squeezing
versus time is plotted in the Figure 6, as is the optimal angle $\theta$
for the spin squeezing. Figure 7 plots the optimal spin squeezing
for a given $N$ in agreement with the plots of Li et al \cite{yunli-2}.
} 
\begin{figure}
\includegraphics[width=0.5\columnwidth]{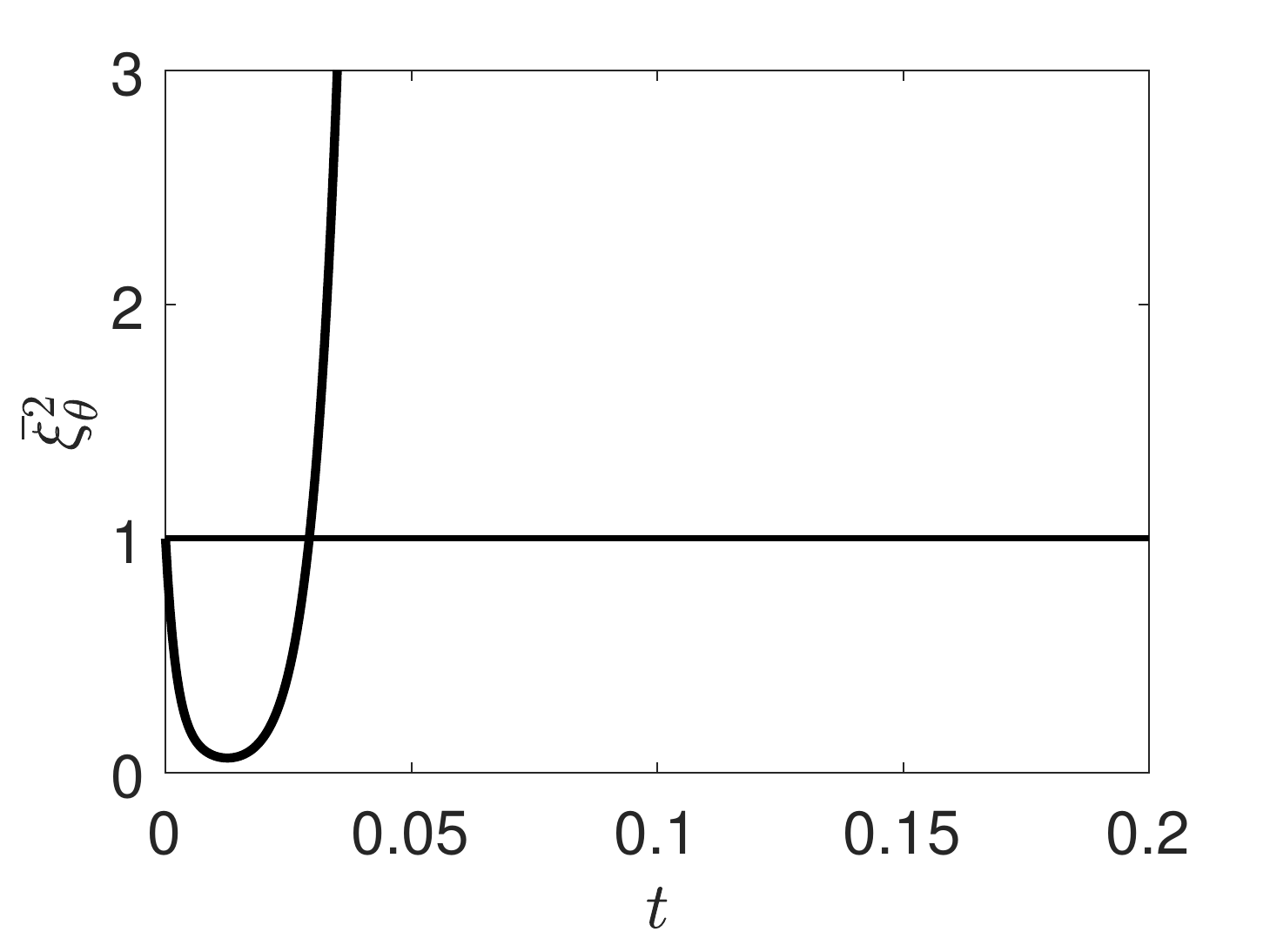}\includegraphics[width=0.5\columnwidth]{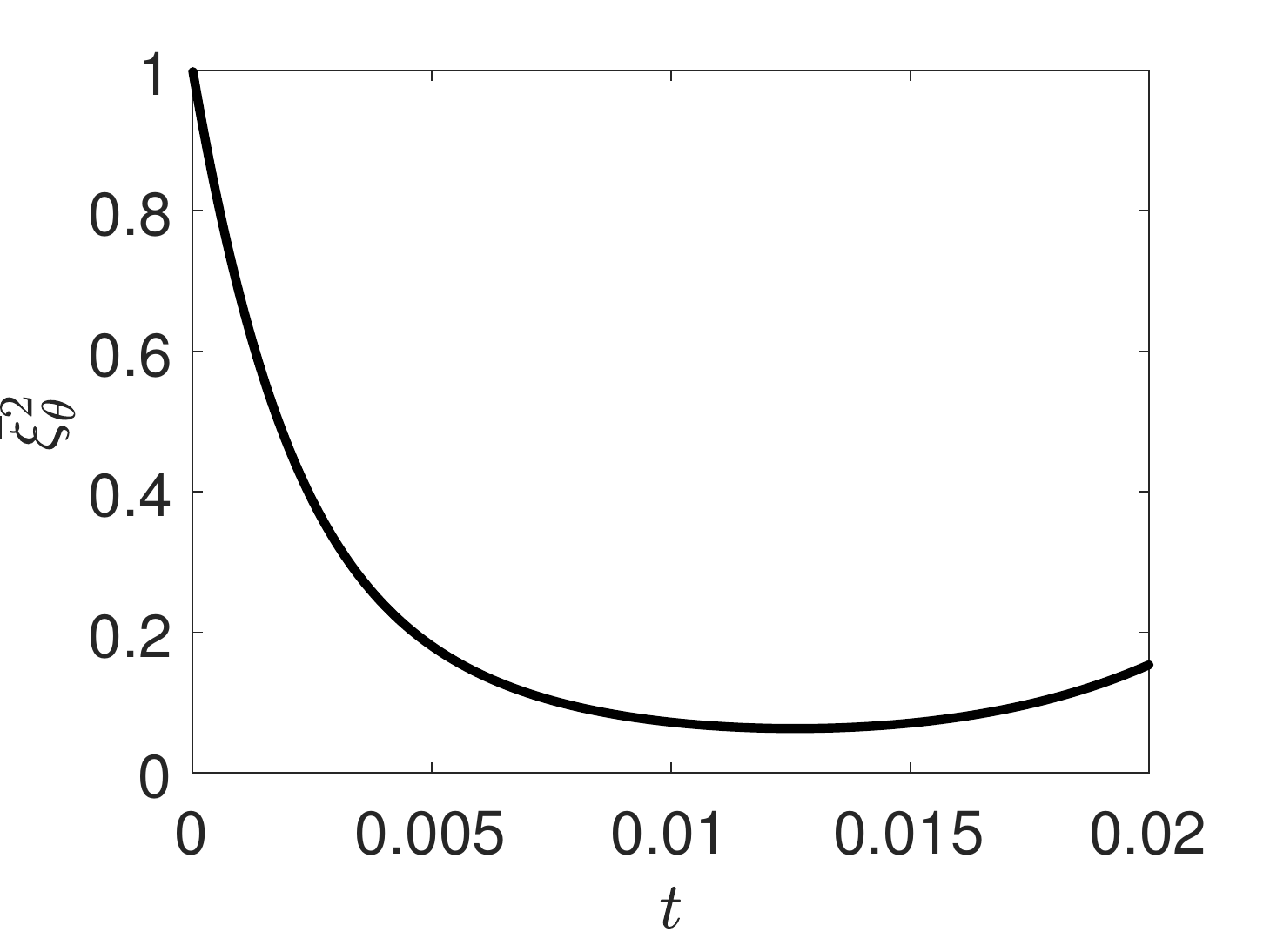}

\includegraphics[width=0.5\columnwidth]{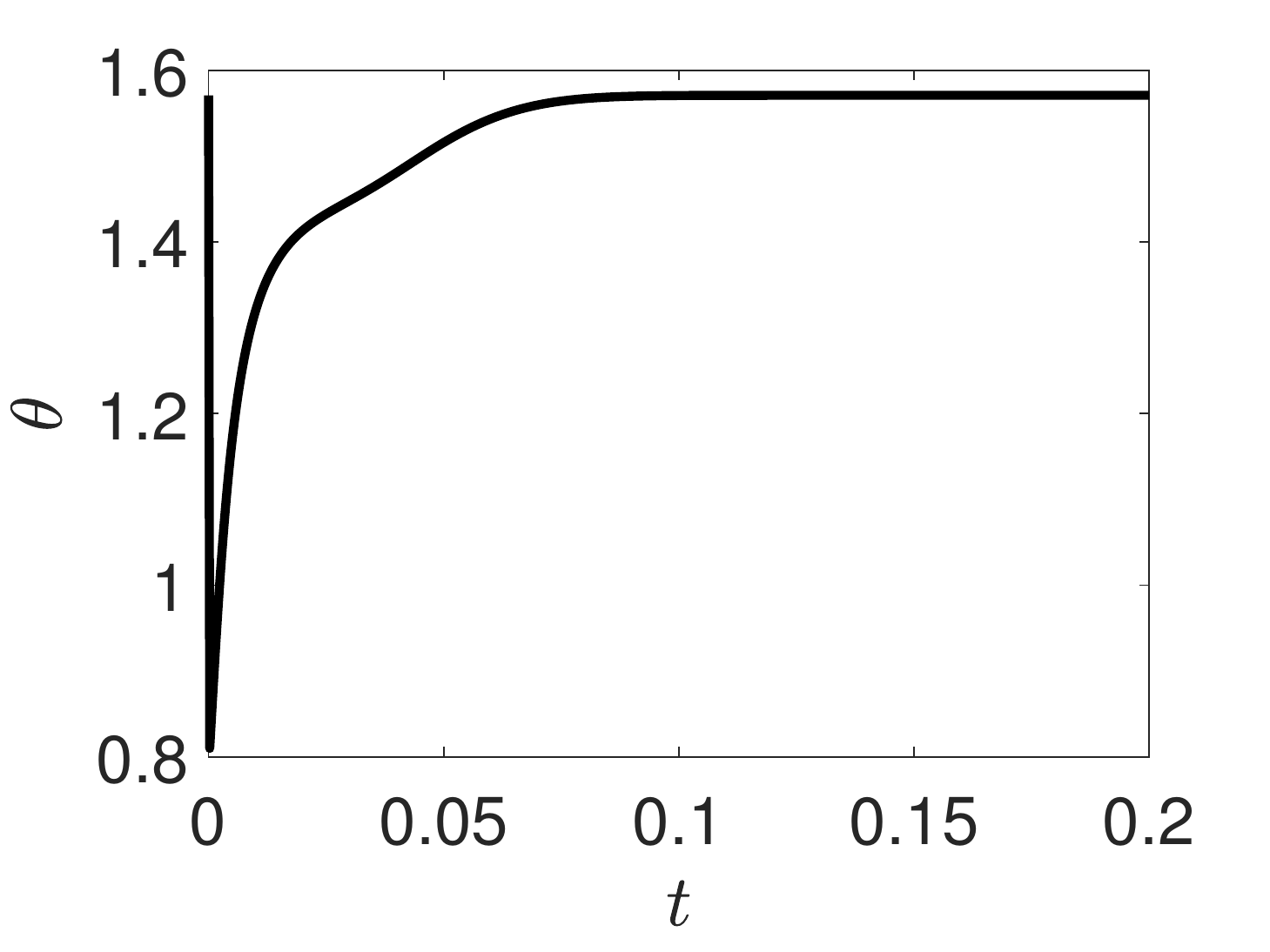}\includegraphics[width=0.5\columnwidth]{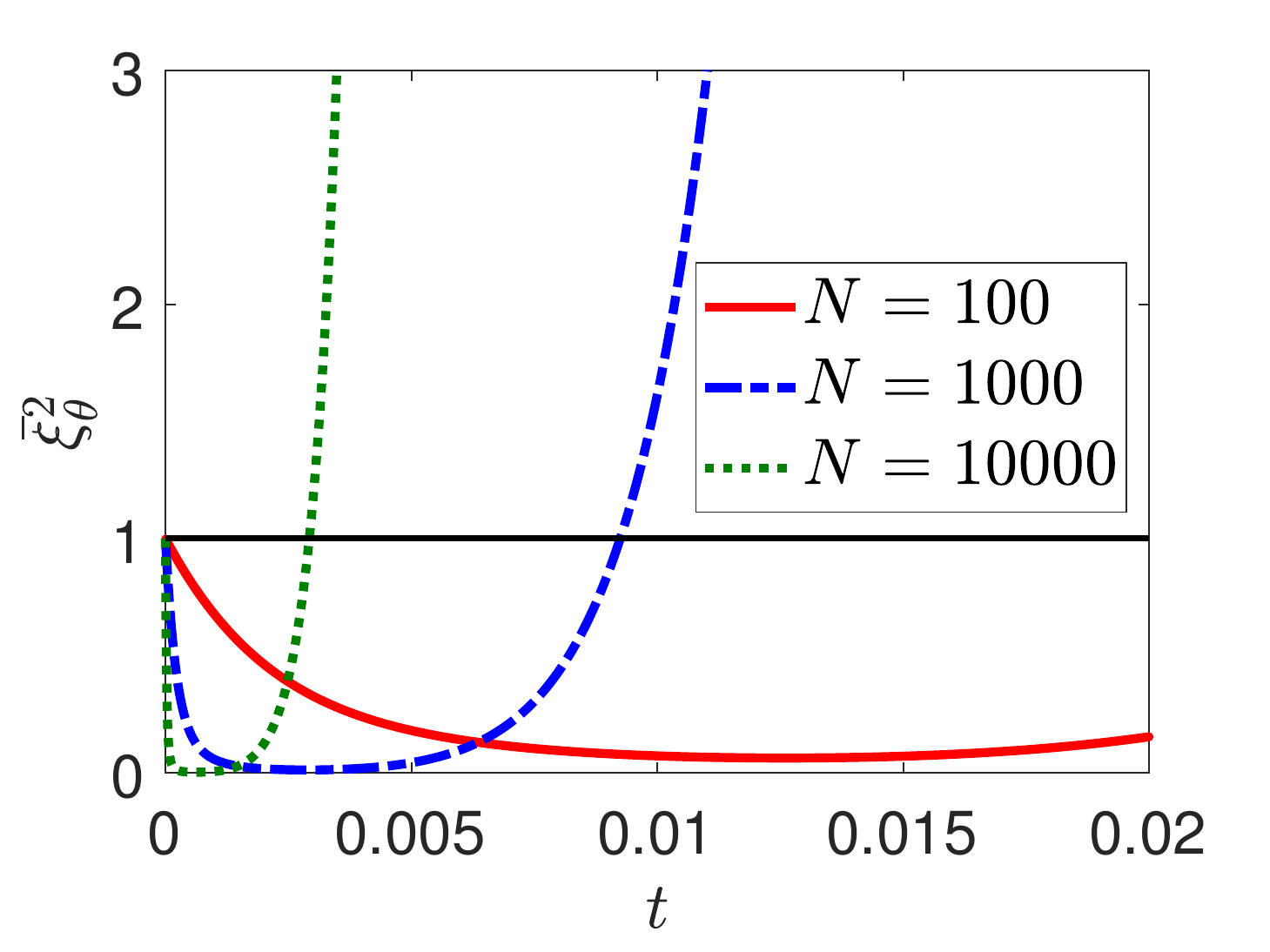}

\caption{\textcolor{black}{The graphs show the optimal spin squeezing for the
fields created in the nonlinear interferometer after evolution for
a time $t$, as in the refs. Li et al. \cite{yun-li,yunli-2}. Here
time $t$ is in units of $1/\chi$. Top figures show the evolution
of spin squeezing as defined by the spin squeezing parameter $\bar{\xi}_{\theta}^{2}$
for $K=-1$ and $N=100$. Spin squeezing is obtained if $\bar{\xi}_{\theta}^{2}<1$.}\textcolor{red}{{}
}\textcolor{black}{The right graph shows the detail over shorter timescales.
The plots of $\xi_{\theta}^{2}$ are indistinguishable from those
of $\bar{\xi}_{\theta}^{2}$ over the timescales for squeezing. The
lower left graph shows the angle $\theta$ for the optimal squeezing
where $N=100$. The lower right graph shows the timescales for higher
atom numbers.}\textcolor{red}{{} }}
\end{figure}

\textcolor{black}{The paper Li et al makes a careful comparison between
the simplistic two-mode model and more complete models that account
for the dynamical changes in the wave function \cite{yunli-2}. They
evaluate $\chi$ for Rb condensates. Figures 2 and 3 of their paper
identify parameter regimes for $N\sim1000$ atoms where the predictions
given by Figure 6 correspond to timescales of order milliseconds and
seconds and are in good agreement with the more accurate models. } 
\begin{figure}
\textcolor{black}{\includegraphics[width=0.5\columnwidth]{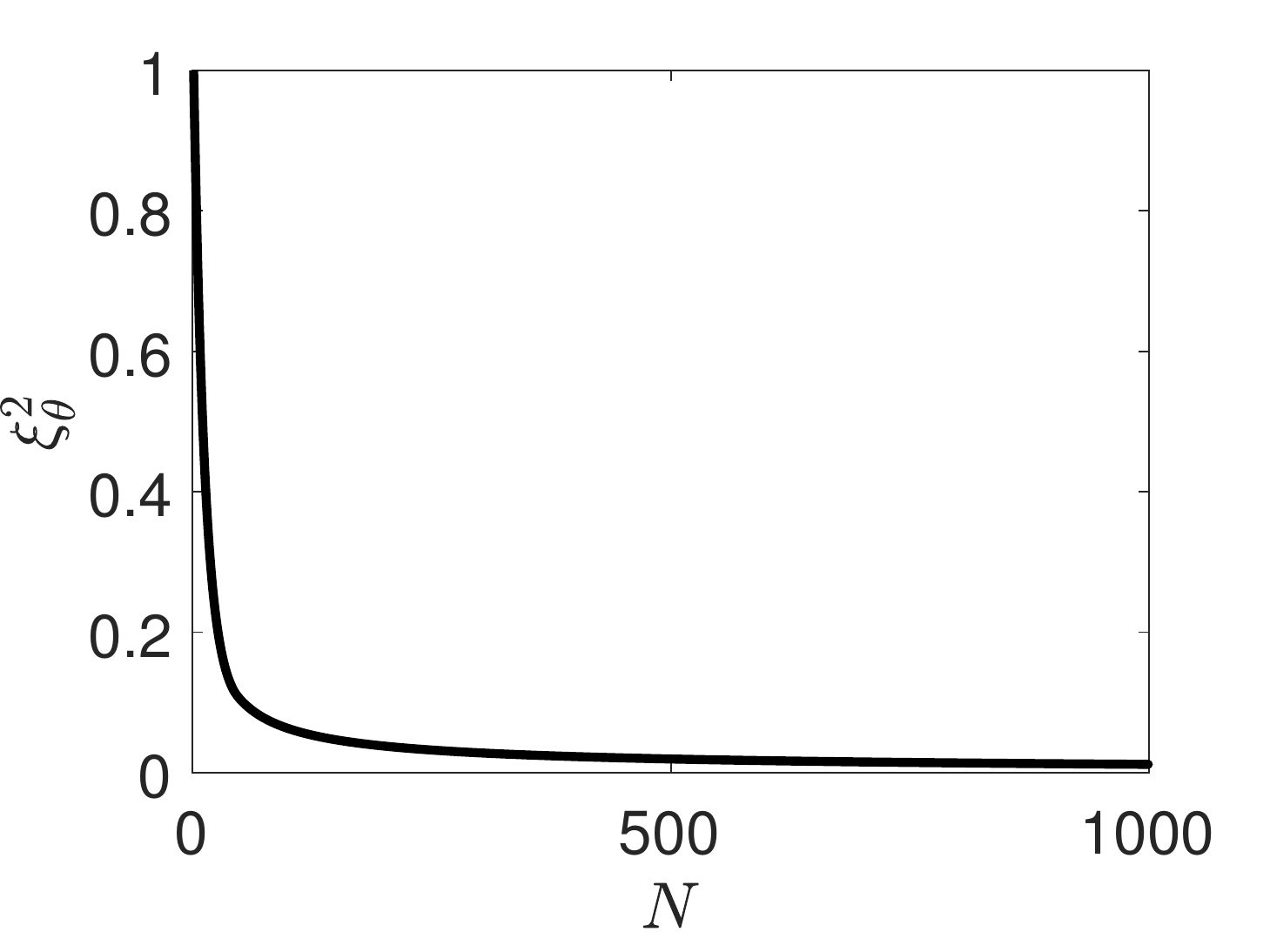}\includegraphics[width=0.5\columnwidth]{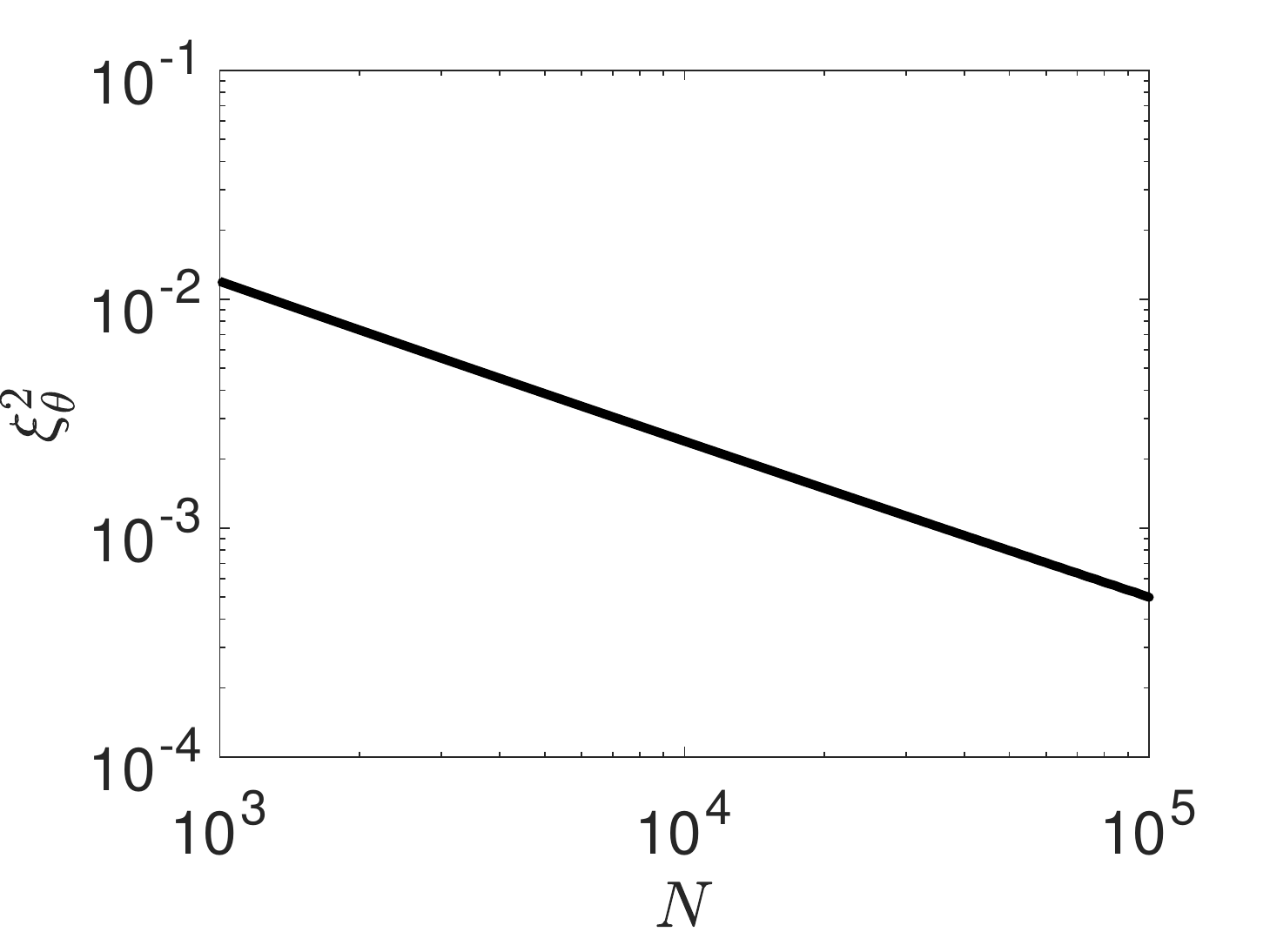}}

\caption{\textcolor{black}{The optimum spin squeezing as defined by the parameter
$\bar{\xi}_{\theta}^{2}$ for each value of $N$. Parameters are as
for Figure 6. The optimisation is done with respect to time $t$ and
angle $\theta$ that defines the squeezing direction.}\textcolor{red}{{} }}
\end{figure}

\subsection{\textcolor{black}{Two-mode EPR steering}}

\begin{figure}
\includegraphics[width=0.5\columnwidth]{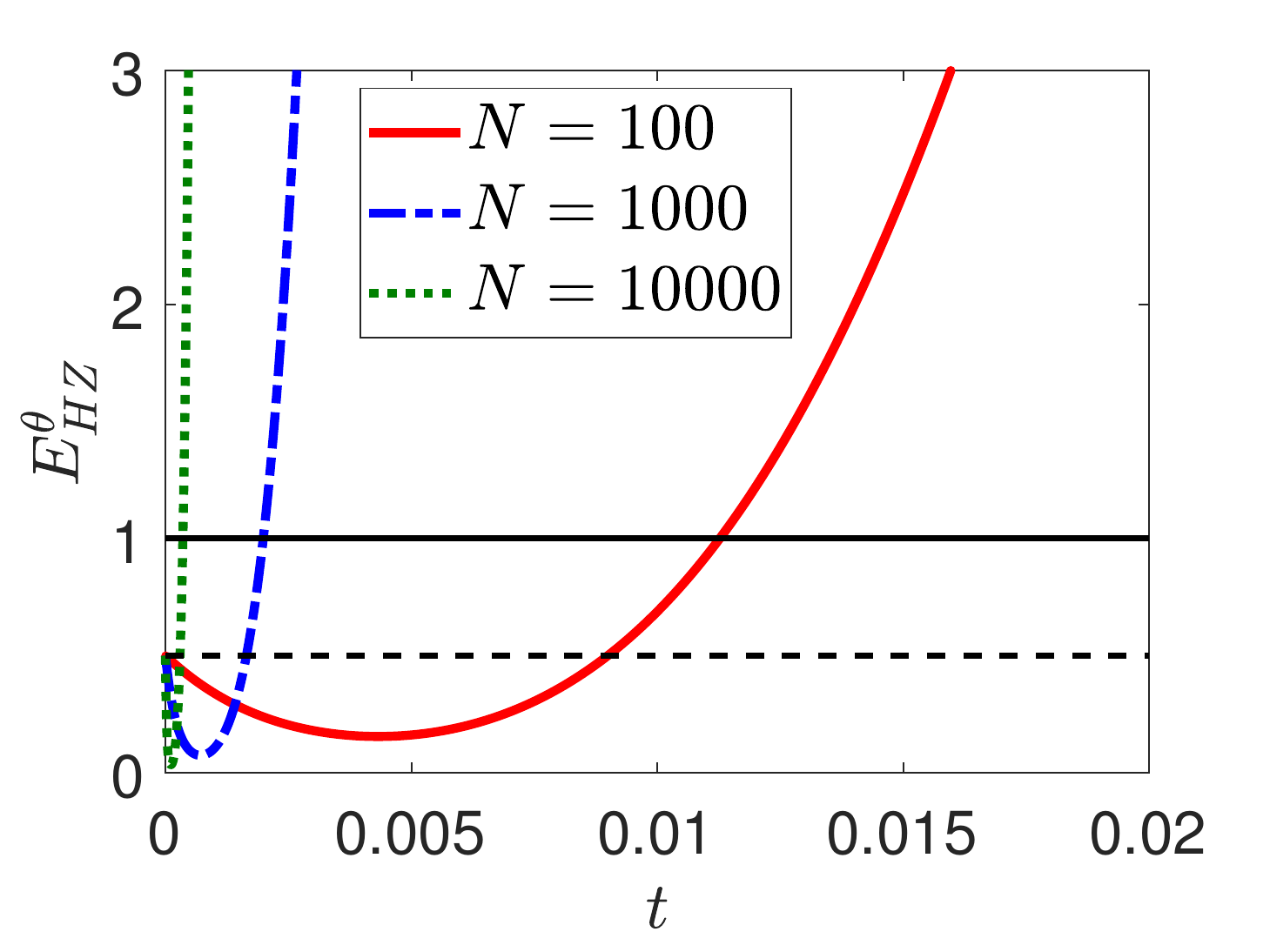}\includegraphics[width=0.5\columnwidth]{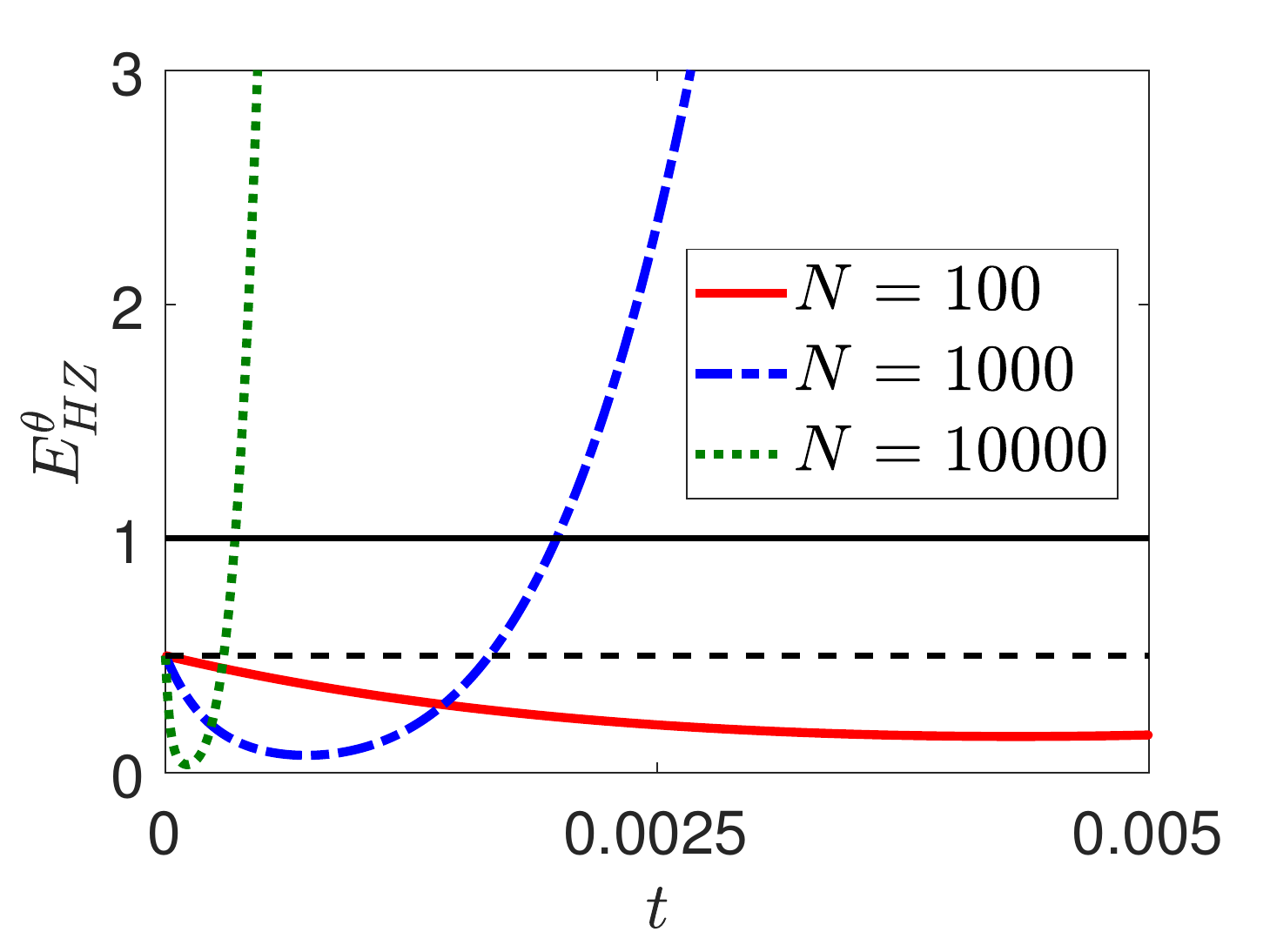}

\caption{\textcolor{black}{The EPR-steering for optimally rotated modes $c_{\theta}$
and $d_{\theta}$ of the nonlinear interferometer after evolution
for a time $t$. The plots show $E_{HZ}^{\theta}$ for the optimal
value of $\theta$ at each time $t$ for $N=100,$ $1000$ and $10^{4}$
atoms. Here $K=-1$ and $t$ is in units of $1/\chi$.}\textcolor{red}{{}
}\textcolor{blue}{{} }\textcolor{black}{Entanglement is signified if
$E_{HZ}^{\theta}<1$. Steering is signified if $E_{HZ}^{\theta}<0.5$.
This implies entanglement and steering between the rotated modes $c_{\theta}$
and $d_{\theta}$ as defined in the text.}}
\end{figure}

\begin{figure}
\includegraphics[width=0.5\columnwidth]{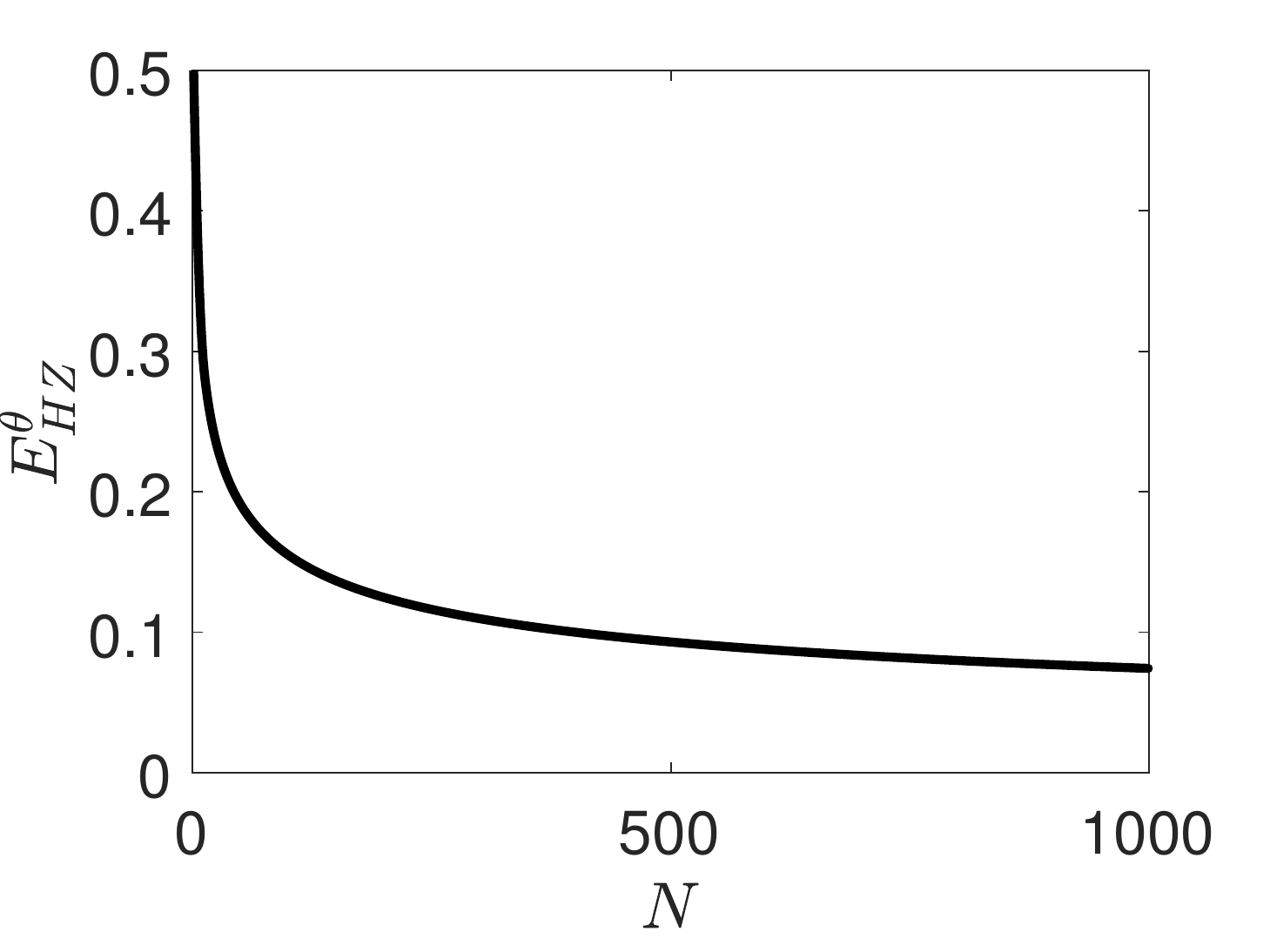}\includegraphics[width=0.5\columnwidth]{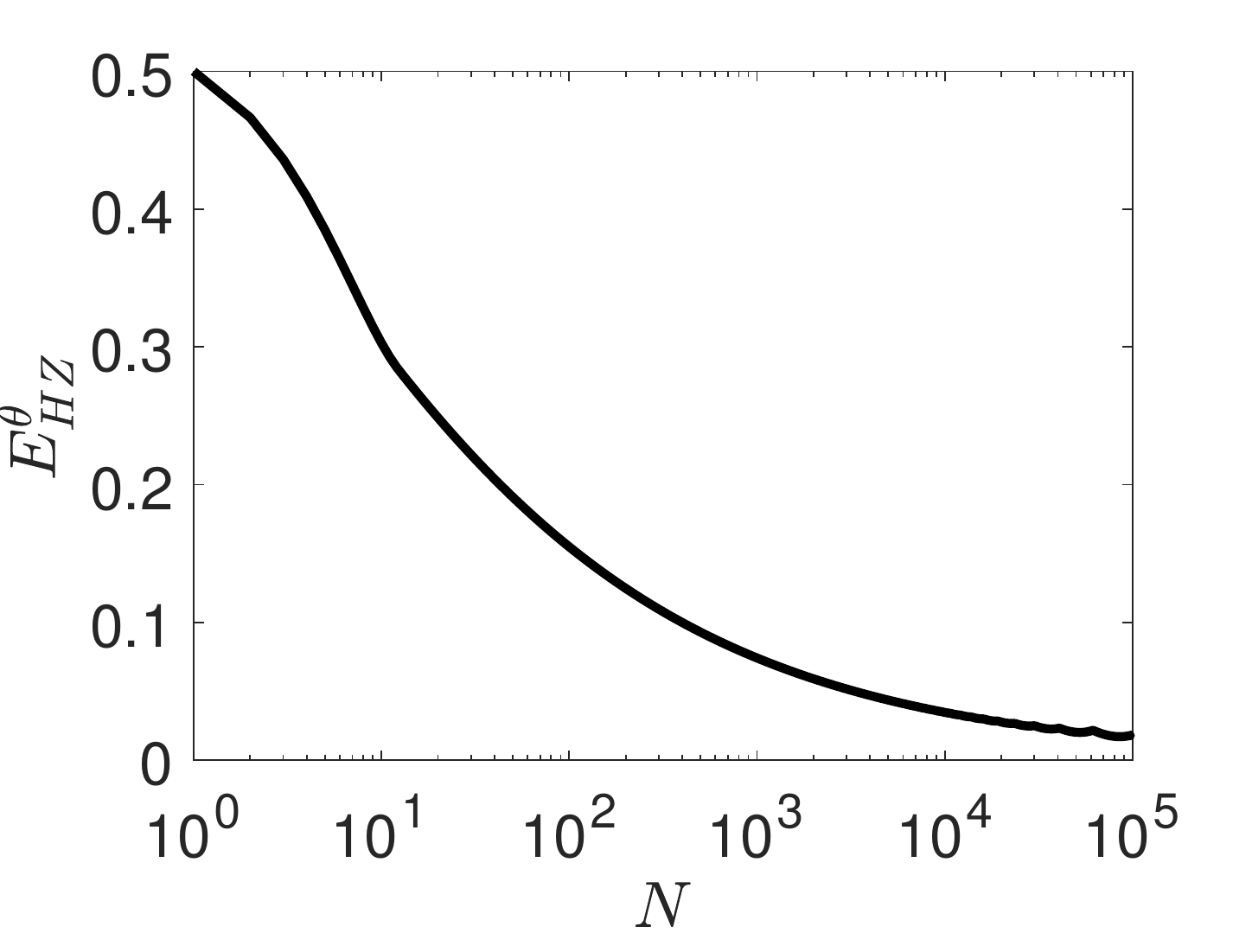}

\includegraphics[width=0.5\columnwidth]{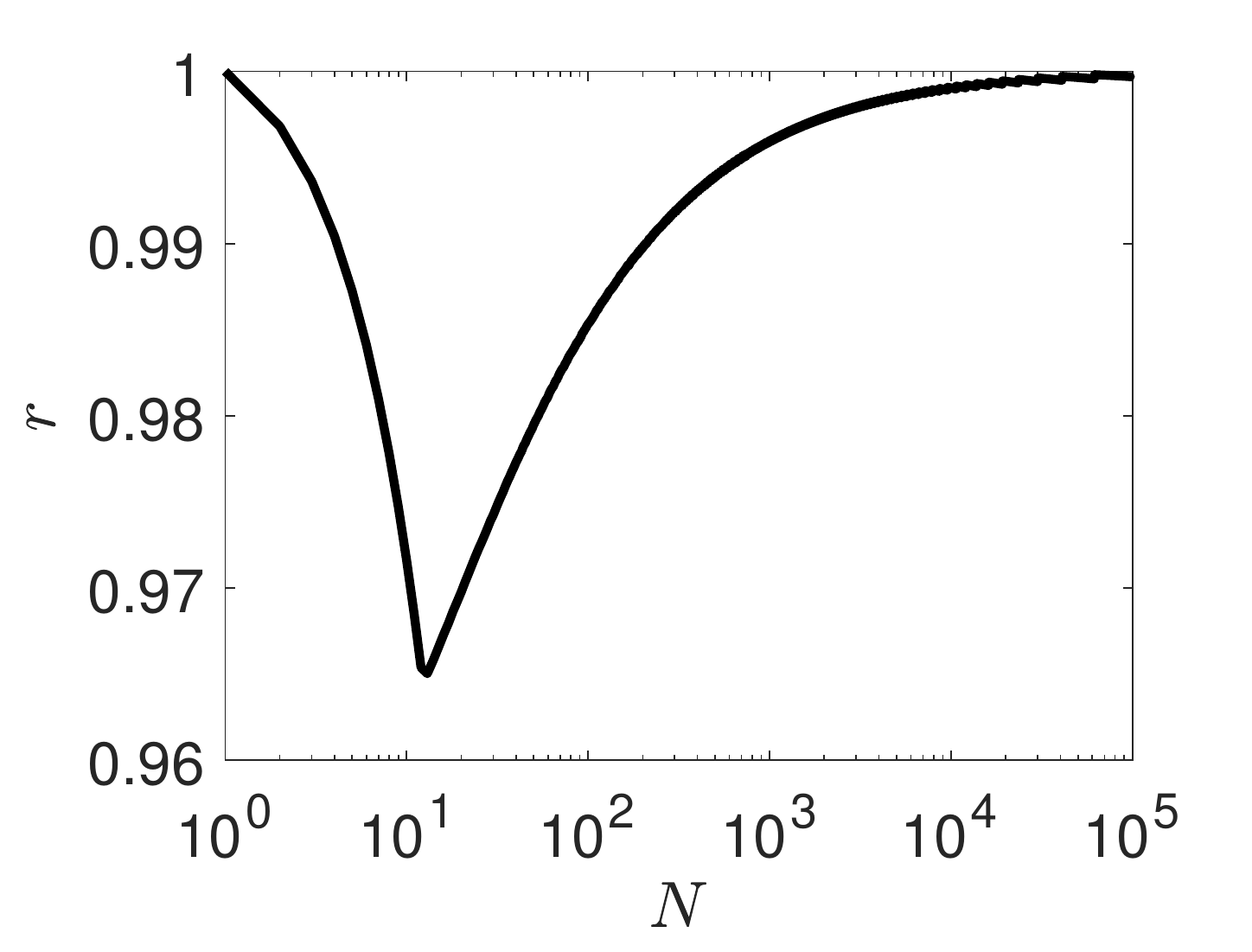}\includegraphics[width=0.5\columnwidth]{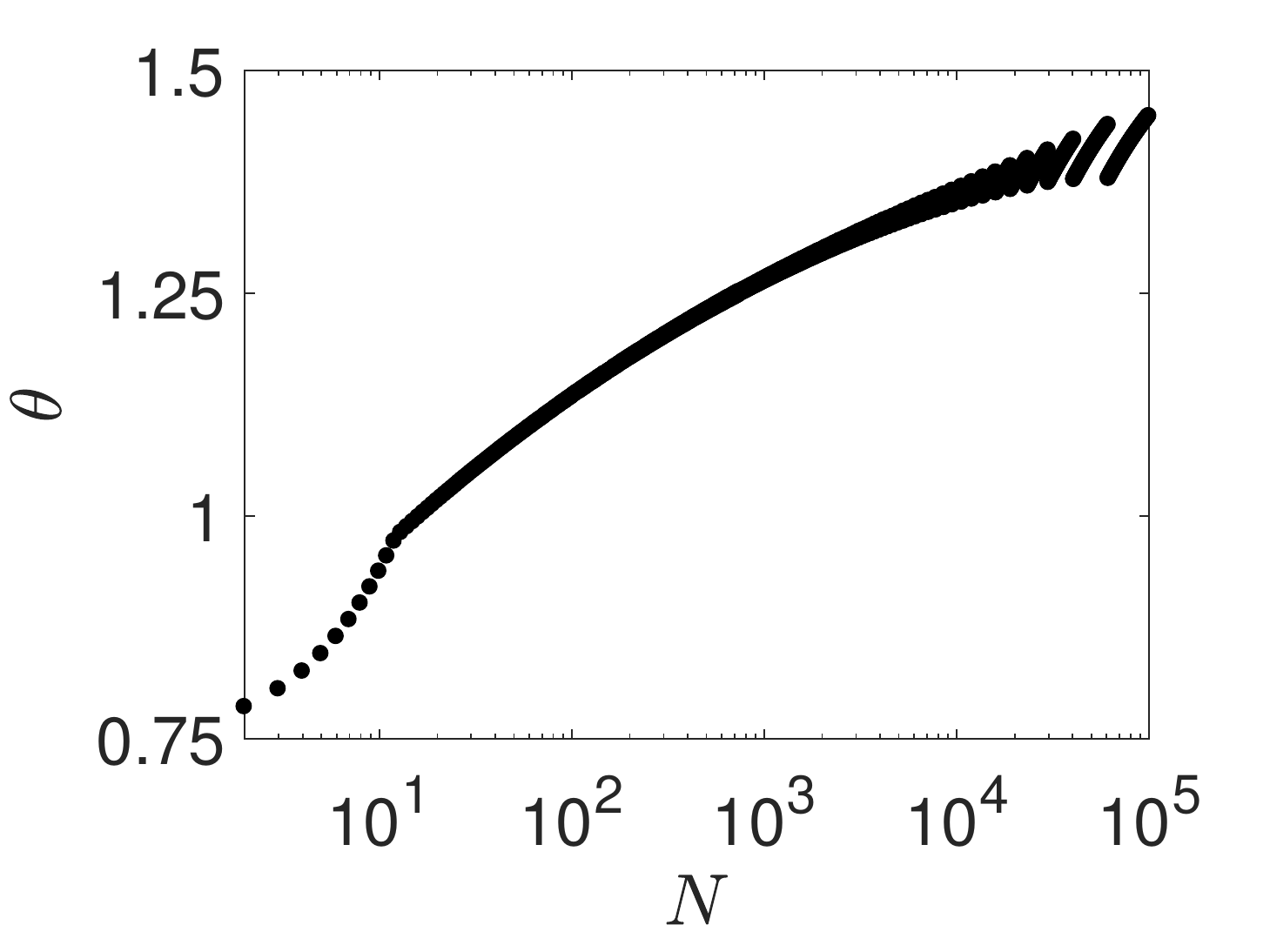}

\caption{\textcolor{black}{The optimum EPR-steering created in the nonlinear
interferometer for a given $N$. Here we optimise with respect to
both time $t$ and angle $\theta$. Figures show the optimal value
$E_{HZ}^{\theta}$ (top) and the corresponding value of $r=\frac{|\langle\vec{S}\rangle|}{\langle\hat{N}\rangle/2}$
(lower left) for those optimised parameters (right). Here $K=-1$.}\textcolor{red}{{}
}\textcolor{black}{Entanglement is signified if $E_{HZ}^{\theta}<1$.
EPR steering is signified if $E_{HZ}^{\theta}<0.5$. The lower right
graph gives the angle $\theta$ that corresponds to the optimal $E_{HZ}^{\theta}$
value for each $N$.}}
\end{figure}

\textcolor{black}{Choosing the direction $\theta$ for optimal squeezing,
we can now define the planar spin variance parameter in the plane
defined by $x$ and $\theta$ as 
\begin{equation}
E_{HZ}^{\theta}=\frac{(\Delta\hat{S}_{x})^{2}+(\Delta\hat{S}_{\theta})^{2}}{\langle\hat{N}\rangle/2}\label{eq:ehztheta}
\end{equation}
The $E_{HZ}^{\theta}$ is plotted in Figure 8. The plots show $E_{HZ}^{\theta}<0.5$
indicating EPR steering (see below). However, the EPR steering signature
implies EPR steering between two modes $c_{\theta}$ and $d_{\theta}$
that are rotated with respect to $a$ and $b$. We need to define
those modes in terms of $a$ and $b$, so that they can spatially
separated in a future experiment that may measure an actual EPR steering.
In fact, the rotated modes are defined according to boson operators
\begin{eqnarray*}
\hat{c}_{\theta} & = & \cos(\theta/2)\hat{a}+i\sin(\theta/2)\hat{b}\\
\hat{d}_{\theta} & = & i\sin(\theta/2)\hat{a}+\cos(\theta/2)\hat{b}
\end{eqnarray*}
 This rotation can be achieved physically by first applying a phase
shifting to mode $a$ by $\pi/2$ so that $\hat{a}\rightarrow i\hat{a}$
followed by a rotation of angle $\theta$, to give new modes $\hat{c}'=i\cos(\theta/2)\hat{a}-\sin(\theta/2)\hat{b}$
and $\hat{d}'=i\sin(\theta/2)\hat{a}+\cos(\theta/2)\hat{b}$. A second
phase shift of $-\pi/2$ is applied to the mode $\hat{c}'$ so that
$\hat{c}'\rightarrow-i\hat{c}'$ and the final transformation is given
by $\hat{c}_{\theta}=\cos(\theta/2)\hat{a}+i\sin(\theta/2)\hat{b}$.
Defining spin operators in the new modes: $\hat{S}_{x}^{\theta}=(\hat{c}_{\theta}^{\dagger}\hat{d}_{\theta}+\hat{c}_{\theta}\hat{d}_{\theta}{}^{\dagger})/2$,
$\hat{S}_{y}^{\theta}=(\hat{c}_{\theta}^{\dagger}\hat{d}_{\theta}-\hat{c}_{\theta}\hat{d}_{\theta}{}^{\dagger})/(2i)$,
$\hat{S}_{z}^{\theta}=(\hat{c}_{\theta}^{\dagger}\hat{c}_{\theta}-\hat{d}_{\theta}{}^{\dagger}\hat{d}_{\theta})/2$,
we find 
\begin{eqnarray}
\hat{S}_{x}^{\theta} & = & \hat{S}_{x}\nonumber \\
\hat{S}_{z}^{\theta} & = & \hat{S}_{z}\cos\theta-\hat{S}_{y}\sin\theta\nonumber \\
\hat{S}_{y}^{\theta} & = & \hat{S}_{z}\sin\theta+\hat{S}_{y}\cos\theta\label{eq:spinrot}
\end{eqnarray}
We note $\hat{S}_{y}^{\theta}=\hat{S}^{\theta}$ where $\hat{S}^{\theta}$
is defined above by Eq. (\ref{eq:spintheta}) and thus $(\Delta\hat{S}_{\theta})^{2}=(\Delta\hat{S}_{y}^{\theta})^{2}$.
Applying the results summarised in Section II for two-mode systems,
we see that entanglement is certified between the modes $c_{\theta}$
and $d_{\theta}$ if one can verify 
\begin{equation}
E_{HZ}^{\theta}<1\label{eq:ehzentspintheta}
\end{equation}
An EPR-steering ($d$ by $c$) is certified if $E_{HZ}^{\theta}<\frac{\langle\hat{c}^{\dagger}\hat{c}\rangle}{\langle\hat{N}\rangle}$
or ($c$ by $d$) if $E_{HZ}^{\theta}<\frac{\langle\hat{d}^{\dagger}\hat{d}\rangle}{\langle\hat{N}\rangle}$.
Since $\langle\hat{c}^{\dagger}\hat{c}\rangle=\langle\hat{N}\rangle/2+\langle\hat{S}_{z}^{\theta}\rangle$
and $\langle\hat{d}^{\dagger}\hat{d}\rangle=\langle\hat{N}\rangle/2-\langle\hat{S}_{z}^{\theta}\rangle$,
these criteria for steering can be rewritten as 
\begin{equation}
E_{HZ}^{\theta}<\frac{1}{2}\pm\frac{\langle\hat{S}_{z}^{\theta}\rangle}{\langle\hat{N}\rangle}\label{eq:ehzspinsteertheta-2}
\end{equation}
EPR steering will always be confirmed in at least one direction if
$E_{HZ}^{\theta}<0.5$. Hence, since the plots of Figure 8 show $E_{HZ}^{\theta}<0.5$,
EPR steering is predicted possible. In fact, we can confirm that in
this case, $\langle\hat{c}^{\dagger}\hat{c}\rangle=\langle\hat{d}^{\dagger}\hat{d}\rangle$
and the observation of $E_{HZ}^{\theta}<0.5$ certifies a two-way
steering. Figure 9 shows the optimal $E_{HZ}^{\theta}$ values for
each $N$. }

\section{Mesoscopic steerable states}

We now apply the criterion developed in Sections II and III, to infer
the depth of two-mode EPR steering. Figures 10-11 give the calibration
showing the effectiveness of the criterion versus $N$, for the steerable
states produced by the nonlinear model. The states generated by the
Hamiltonian are pure steerable states with a total number of atoms
given by $N$. The criterion is used to place a rigorous \emph{lower}
bound based only on the observed experimental variances, without the
assumption of a pure state. However a maximally effective criterion
would detect a depth of steering of $n_{st}\sim N$. This value is
not detected with the criteria, because the states of the interferometer
are not the maximal planar spin squeezed states \cite{cj-2}. Summarising
the Result 2 proved in Section III.B, if we measure $\frac{E_{HZ}}{r}<\tilde{C}_{s_{0}}$
where $\tilde{C}_{s_{0}}=C_{s_{0}}/s_{0}$ and $r=\frac{|\langle\vec{S}\rangle|}{\langle\hat{N}\rangle/2}$,
we deduce a depth of EPR steering of (at least) $n_{st}\sim2s_{0}$.

\begin{figure}
\includegraphics[width=0.6\columnwidth]{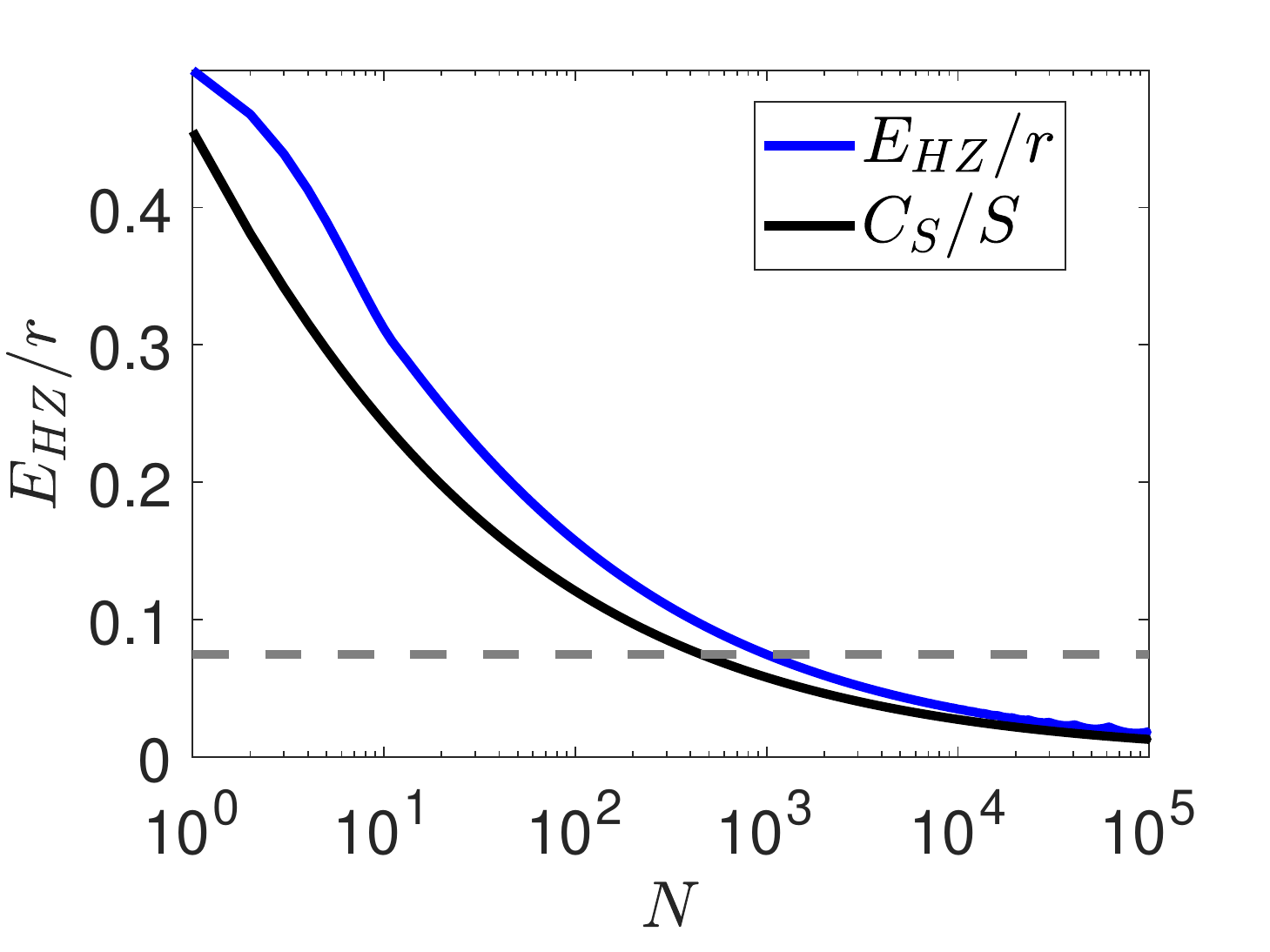}

\caption{The value of $\frac{E_{HZ}^{\theta}}{r}$ for the interferometer with
$N$ atoms is shown by the upper blue line. Here we have evaluated
$\frac{E_{HZ}^{\theta}}{r}$ using the angle $\theta$ and time $t$
that minimises the value of $E_{HZ}^{\theta}$ for a given $N$, as
displayed in Figure 9. Here $K=-1$. This value may be compared with
the fundamental lower value given by the plot of \textcolor{black}{$\tilde{C}_{S}=C_{S}/S$
with $S=N/2$ (lower black line).}}
\end{figure}

\begin{figure}
\includegraphics[width=0.5\columnwidth]{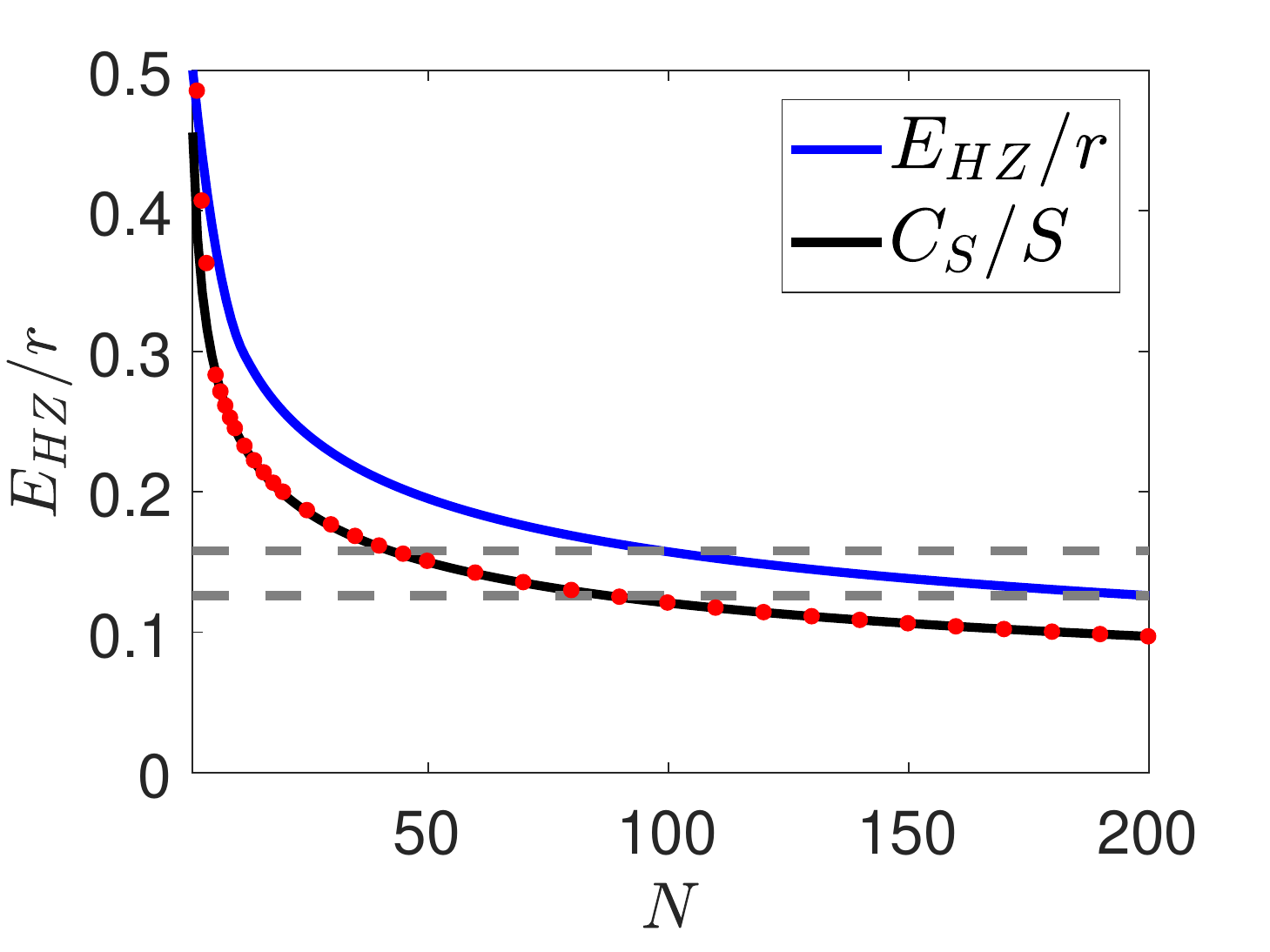}\includegraphics[width=0.5\columnwidth]{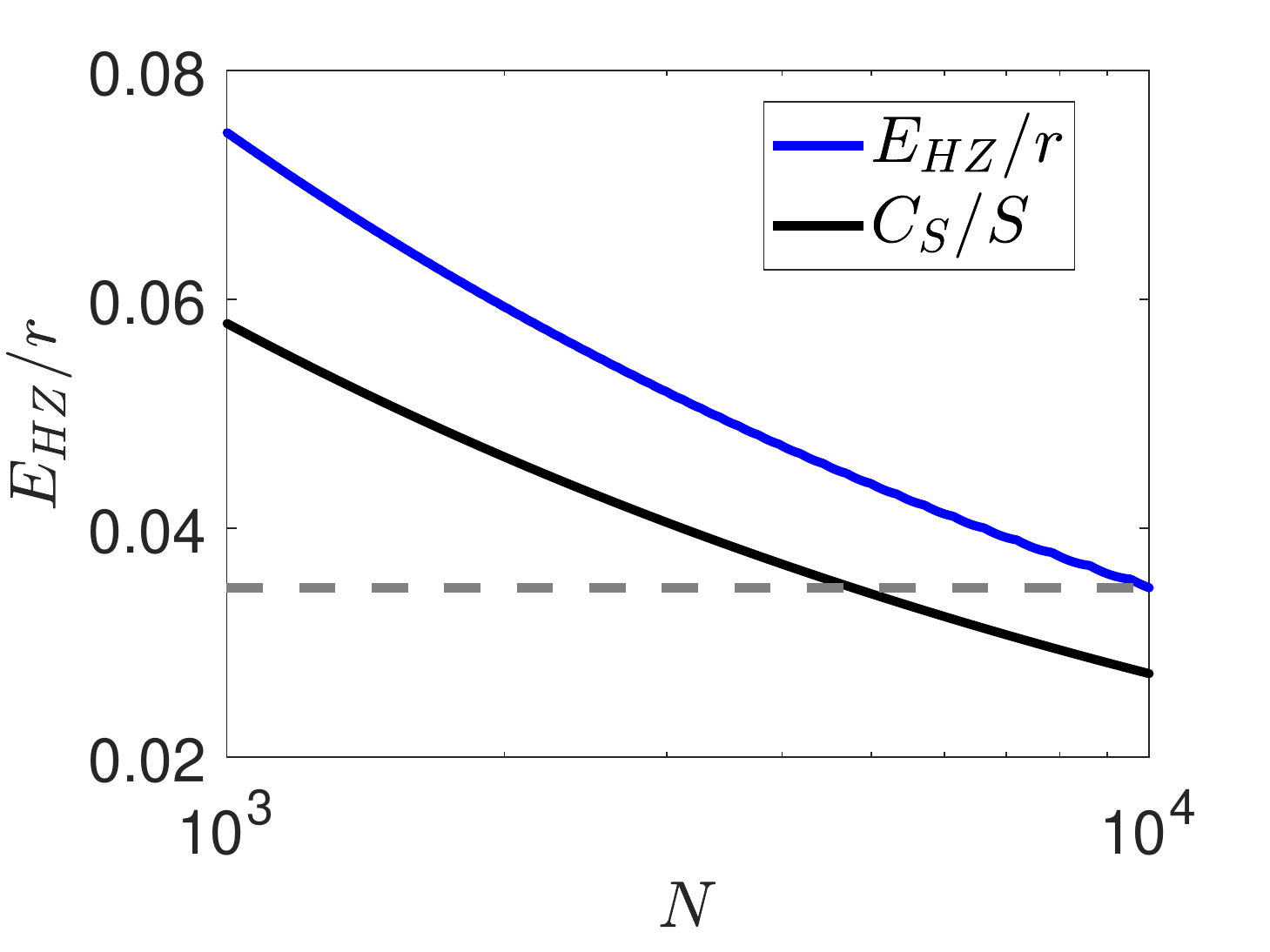}

\caption{Curves are as in Figure 10. \textcolor{black}{For the left graph,
the top horizontal dashed line cuts the $C_{S}/S$ curve just below
$C_{S}/S=0.1581$. We see that}\textcolor{blue}{{} }\textcolor{black}{this
value corresponds to $N=2S=42$. Hence if the value $E_{Hz}/r$ is
indeed measured, we deduce a steerable state with $n_{st}=42$ atoms.
A similar line is drawn for $N=200$ }\textcolor{blue}{{} }\textcolor{black}{The
right graph show the calibration for higher $N$.}}
\end{figure}

In the Figures 10 -12 we plot the predictions for $\frac{E_{HZ}^{\theta}}{r}$
versus $N$ that are shown in Figure 9 for the two-mode BEC atom interferometer.
\textcolor{black}{The graphs also show the calibration curves based
on the results for $C_{S}/S$ as given in Ref. \cite{cj-2} where
we have put $N=2S$. These curves extend Figure 1 to larger $S$ and
give the fundamental lower bound of the Hillery-Zubairy planar spin
squeezing parameter $E_{HZ}/r$. This fundamental lower bound is determined
by quantum mechanics. Figure 10 plots the comparison for large values
of $N$ based on the analytical expressions given in the paper He
et al \cite{cj-2}. For high $N$ the lines become indistinguishable
on the linear scale for the variances. } 
\begin{table}
\begin{centering}
\begin{tabular}{|c|c|c|c|}
\hline 
\textcolor{black}{$N$}  & \textcolor{black}{$\frac{E_{HZ}^{\theta}}{r}$}  & \textcolor{black}{${\color{red}{\color{black}2}}S$}  & \textcolor{black}{$\tilde{C}_{S}=\frac{C_{S}}{S}$}\tabularnewline
\hline 
\textcolor{black}{$50$}  & \textcolor{black}{$0.1951$}  & \textcolor{black}{$21$}  & \textcolor{black}{$0.1952$}\tabularnewline
\hline 
\textcolor{black}{$100$}  & \textcolor{black}{$0.1572$}  & \textcolor{black}{$42$}  & \textcolor{black}{$0.1581$}\tabularnewline
\hline 
\textcolor{black}{$200$}  & \textcolor{black}{$0.1261$}  & \textcolor{black}{$87$}  & \textcolor{black}{$0.1262$}\tabularnewline
\hline 
\textcolor{black}{$500$}  & \textcolor{black}{$0.0936$}  & \textcolor{black}{$223$}  & \textcolor{black}{$0.0938$}\tabularnewline
\hline 
\textcolor{black}{$1000$}  & \textcolor{black}{$0.07457$}  & \textcolor{black}{$456$}  & \textcolor{black}{$0.07459$}\tabularnewline
\hline 
\textcolor{black}{$10000$}  & \textcolor{black}{$0.034775$}  & \textcolor{black}{$4772$}  & \textcolor{black}{$0.034776$}\tabularnewline
\hline 
\end{tabular}
\par\end{centering}
\textcolor{black}{\caption{\textcolor{black}{Value of the ratio $\frac{E_{HZ}^{\theta}}{r}$
for different values of $N$. Values of $C_{S}$ for different values
of $S$ using the analytical expressions given in the paper He et
al \cite{cj-2} that corresponds to the minimum value where the condition
$\frac{E_{HZ}}{r}<\tilde{C}_{S}$ is satisfied.}}
} 
\end{table}

\textcolor{black}{Figure 11 gives the close-up of the predictions
of $\frac{E_{HZ}^{\theta}}{r}$ for the BEC intereferometer for $N\sim100-200$
atoms.} For a given value of $N$, the predictions may be compared
with the plot of \textcolor{black}{$\tilde{C}_{S}=C_{S}/S$ with $N=2S$
given by the lower black line. At $N=100$, we see that $E_{HZ}^{\theta}/r=0.1572$.
The horizontal grey dashed line on the graph gives for $N=100$ the
minimum value $\tilde{C}_{S}=0.1581$ satisfying the condition (\ref{eq:proc}).
We see that}\textcolor{blue}{{} }\textcolor{black}{this corresponds
to $N=2S=42$ on the calibration curve. Hence if the value $E_{HZ}^{\theta}/r$
is indeed measured, one can infer an EPR steerable state with at least
$n_{st}=42$ atoms. A similar line is drawn for $N=200$ indicating
$n_{st}\geq87$. The right graph of Figure 11 gives the same analysis
but for $N=1000-10,000$ atoms. If the predicted amount of planar
spin squeezing $E_{HZ}$ can be observed at $N=10,000$, then it would
be possible to deduce mesoscopic steerable states with thousands of
atoms. The link between the observed value $\frac{E_{HZ}^{\theta}}{r}$
and the number of atoms that can be inferred is given in the Table
I for the range of $N$ that are typical of the spin squeezing experiments.
 }

\section{Discussion }

There are two main assumptions of the theory given in Section IV of
this paper\textcolor{black}{. First, the interferometer is modelled
using a simple two-mode Hamiltonian that ignores loss of atoms into
other modes, and also ignores the spatial dynamics of the mode functions.
More complete treatments show the validity of the approximation, at
least for calculations of the spin squeezing, over certain regimes
\cite{yunli-2,bogdanepl,asinatra}. A treatment allowing for changes
in both mode occupancy and the mode functions is set out in Ref. \cite{bjdtwo-mode}.
Li et al give a detailed comparison between the two-mode model and
more complete models that account for the spatial dynamics. The predictions
of the simple model are found to be achievable for Rb condensates
of $N\sim10^{4}$ atoms \cite{yunli-2}. The conclusion of the present
paper is that one can expect an evolution of the EPR steering correlations
over similar timescales as the evolution of spin squeezing. The full
calculation however involves the dynamics of the variance associated
with the Bloch vector, which was not studied earlier.}

\textcolor{black}{At higher temperatures and for larger numbers of
atoms, a multi-mode model will be necessary. EPR steering correlations
are known to be sensitive to thermal noise, which has not been included
in this paper \cite{thermal}. A multimode treatment that fully accounts
for spatial variation of the wavefunctions has been given by Opanchuk
et al \cite{bogdanepl}. The depth of EPR steering criterion used
in this paper gives a lower bound of the number of atoms in the }\textcolor{black}{\emph{two-mode}}\textcolor{black}{{}
steerable state. Where other modes are present, we point out it is
possible to extract the relevant two-mode condensate moments from
experimental data, so that the criteria can be applied. This will
be discussed in another paper. }

\textcolor{black}{The second assumption made in the theory is that
there is a fixed number $N$ of atoms incident on the interferometer.
While this is typical of many models that have successfully described
the evolution of spin squeezing, in practice the fluctuating number
input and the inability to fully resolve the atom number on measurement
will introduce deviations from the theory. Current experimental strategies
do not allow precise control of the number inputs. The effect of number
fluctuations on the Hillery-Zubairy entanglement parameter has been
studied in the Refs. \cite{timpaper-2,bogdaneprbec}. The variance
of $\hat{S}_{x}$ (the Bloch vector) is directly related to the variance
of the number input and will increase with increased number fluctuations.
Yet, a reduction in the variance for $\hat{S}_{x}$ below the Poissonian
level is required for the Hillery-Zubairy EPR steering signature.
However, it was also shown in the papers \cite{timpaper-2,bogdaneprbec}
that the Hillery-Zubairy criterion can be modified by a normalisation
with respect to total number, to give greater sensitivity in the presence
of number fluctuations. If the total number of atoms can be accurately
counted on detection, then postselection of states with definite $N$
is possible.}

\section{Conclusion}

In summary, we have analysed theory for a two-mode nonlinear interferometer
and shown that entanglement and EPR steering correlations between
the two modes are predicted. The correlations may be signified by
measuring noise reduction in the sum of two spin variances (planar
spin squeezing) in accordance with a two-mode Hillery-Zubairy parameter.
The required moments are measurable as the population differences
at the output of the interferometer, after appropriate phase shifts
and re-combinations of the modes. These interactions are realised
in atom interferometers as Rabi rotations using microwave pulses.

In principle, it is possible to spatially separate the two modes that
show the EPR steering correlations. It is also possible in principle
to measure the steering correlations by performing local measurements.
This can be seen by expanding the two-mode moments of eqs. (\ref{eq:hzab})
and (\ref{eq:hzsteerab}) in terms of the local quadrature phase amplitudes.
However, the proposal of this paper is to give preliminary evidence
of the EPR steering correlations, by recombining the modes at the
final beam splitter of the interferometer.

Recent experiments report detection of EPR steering for Bose-Einstein
condensates, including for spatial separations. The purpose of our
work is to provide a method to extend such analyses, to quantify the
number of atoms genuinely comprising the EPR steerable state. The
method we give here is based on the lower bounds derived in Ref. \cite{cj-2}
for an uncertainty relation involving two spins, and would be useful
where the steering is identified via planar spin squeezing, or the
Hillery-Zubairy parameter. 
\begin{acknowledgments}
This work has been supported by the Australian Research Council under
Grant DP140104584. We thank Yun Li, B. Opanchuk and P. Drummond for
useful discussions. This work was performed in part at Aspen Center
for Physics, which is supported by National Science Foundation grant
PHY-1607611. BD thanks the Centre for Cold Matter, Imperial College
for hospitality during this research. 
\end{acknowledgments}

\appendix
%dummy comment inserted by tex2lyx to ensure that this paragraph is not empty

\section{Proof of two-mode depth of steering criterion\label{sec:ApAdepth}}

\textbf{\textcolor{black}{Proof:}}\textcolor{black}{{} We follow
the steps and definitions of the previous proof III.B to arrive at
the inequality $(\Delta\hat{S}_{y})^{2}+(\Delta\hat{S}_{z})^{2}\geq\sum_{R}P_{R}\{(\Delta_{R}\hat{S}_{y})^{2}+(\Delta_{R}\hat{S}_{z})^{2}\}$.
We find, using (\ref{eq:planefuns-2}) and the result for all non-steerable
states 
\begin{eqnarray}
(\Delta\hat{S}_{y})^{2}+(\Delta\hat{S}_{z})^{2} & \geq & P_{lhs}0.5\sum_{R''}P_{R''}s_{R''}\nonumber \\
 &  & +P_{st}\sum_{R'}P_{R'}s_{R'}F_{s_{R'}}\Bigl(\frac{\langle S_{||}\rangle_{R'}}{s_{R'}}\Bigr)\nonumber \\
\label{eq:6-1}
\end{eqnarray}
where here we let $s_{R}=\langle\hat{N}\rangle_{R}/2$. Using that
the functions $F_{S}\Bigl(\frac{\langle S_{||}\rangle}{S}\Bigr)$
are monotonically decreasing with $S,$ for fixed $\frac{\langle S_{||}\rangle_{R'}}{s_{R'}}$,
it follows that 
\begin{eqnarray}
(\Delta\hat{S}_{y})^{2}+(\Delta\hat{S}_{z})^{2} & \geq & P_{lhs}0.5\sum_{R''}P_{R''}s_{R''}\nonumber \\
 &  & +P_{st}\sum_{R'}P_{R'}s_{R'}F_{s_{0}}\Bigl(\frac{\langle S_{||}\rangle_{R'}}{s_{R'}}\Bigr)\nonumber \\
\label{eq:7-1}
\end{eqnarray}
where $s_{0}$ is defined in III.B, as the maximum value of the spins
of the set of steerable states. Following the proofs of Refs. \cite{sm-2},
we use that the functions $F_{S}$ are convex. Hence they satisfy
the inequality $\sum_{k}c_{k}F(x_{k})\geq F(\sum_{k}c_{k}x_{k})$
\cite{sm-2} where $c_{k}$ are real numbers. Thus, on introducing
$k=\langle\hat{N}\rangle/2$ 
\begin{eqnarray}
P_{st}\sum_{R'}P_{R'}s_{R'}F_{s_{0}}\Bigl(\frac{\langle S_{||}\rangle_{R'}}{s_{R'}}\Bigr) & = & P_{st}k\sum_{R'}\frac{P_{R'}s_{R'}}{k}F_{s_{0}}\Bigl(\frac{\langle S_{||}\rangle_{R'}}{s_{R'}}\Bigr)\nonumber \\
 & \geq & kF_{s_{0}}\Bigl(\sum_{R'}P_{st}P_{R'}\frac{\langle S_{||}\rangle_{R'}}{k}\Bigr)\nonumber \\
\label{eq:8-1}
\end{eqnarray}
Thus 
\begin{eqnarray}
(\Delta\hat{S}_{y})^{2}+(\Delta\hat{S}_{z})^{2} & \geq & k\Bigl(\sum_{R''}P_{lhs}P_{R''}\frac{s_{R''}}{k}0.5\Bigr)\nonumber \\
 &  & +kF_{s_{0}}\Bigl(\sum_{R'}P_{st}P_{R'}\frac{\langle S_{||}\rangle_{R'}}{k}\Bigr)\nonumber \\
\label{eq:13-1}
\end{eqnarray}
We have taken the case where it is true that $F_{S}\Bigl(x\Bigr)\leq0.5$
for all $x$ \cite{toth-planar-ent}. Thus}

\textcolor{black}{{} 
\begin{eqnarray}
(\Delta\hat{S}_{y})^{2}+(\Delta\hat{S}_{z})^{2} & \geq & k\Bigl(\sum_{R''}P_{lhs}P_{R''}\frac{s_{R''}}{k}F_{s_{0}}(\frac{\langle S_{||}\rangle_{R''}}{s_{R''}}\Bigr))\nonumber \\
 &  & +kF_{s_{0}}\Bigl(\sum_{R'}P_{st}P_{R'}\frac{\langle S_{||}\rangle_{R'}}{k}\Bigr)\nonumber \\
\label{eq:13-1-1}
\end{eqnarray}
Using convexity, we find}

\textcolor{black}{{} 
\begin{eqnarray}
(\Delta\hat{S}_{y})^{2}+(\Delta\hat{S}_{z})^{2} & \geq & kF_{s_{0}}(\sum_{R''}P_{lhs}P_{R''}\frac{\langle S_{||}\rangle_{R''}}{k}\Bigr)\nonumber \\
 &  & +kF_{s_{0}}\Bigl(\sum_{R'}P_{st}P_{R'}\frac{\langle S_{||}\rangle_{R'}}{k}\Bigr)\nonumber \\
 & \geq & kF_{s_{0}}\Bigl(\sum_{R}P_{st}P_{R}\frac{\langle S_{||}\rangle_{R}}{k}\Bigr)\nonumber \\
 & = & \frac{\langle\hat{N}\rangle}{2}F_{s_{0}}\Bigl(\frac{\langle S_{||}\rangle}{\langle\hat{N}\rangle/2}\Bigr)\label{eq:9-1-1}
\end{eqnarray}
The last line can be rewritten 
\begin{equation}
E_{HZ}^{yz}\geq F_{s_{0}}\Bigl(\frac{\langle S_{||}\rangle}{\langle\hat{N}\rangle/2}\Bigr)\label{eq:steerdepth-1}
\end{equation}
Violation of this inequality implies the existence of a steerable
state with spin $s>s_{0}$, which implies a state with greater than
$2s_{0}$ atoms. }

\section{Evaluation of moments \label{sec:ApAEvMoments-1}}

Here, we show the expressions for the moments evaluated from the Hamiltonian
$H_{NL}$ given in Eq. (\ref{eq:fullnlk}). \textcolor{black}{Using
the state $|\psi(t)\rangle$ we evaluate the moments needed in the
expressions for the variances, $E_{HZ}$ and spin squeezing of Section
IV. } 
\begin{eqnarray*}
\langle\hat{a}^{\dagger}\hat{b}\rangle & = & \sum_{k=0}\sum_{r=0}c_{k}^{*}c_{r}e^{i\left(\Omega(k)-\Omega(r)\right)t/\hbar}\sqrt{N-r+1}\sqrt{r}\times\\
 &  & \langle N-k|N-r+1\rangle_{a}\langle k|r-1\rangle_{b}\\
 & = & \sum_{k=0}c_{k}^{*}c_{k+1}e^{i[\Omega(k)-\Omega(k+1)]t/\hbar}\sqrt{N-k}\sqrt{k+1}
\end{eqnarray*}

\begin{eqnarray*}
\langle\hat{a}^{\dagger}\hat{a}\rangle & = & \langle\hat{b}^{\dagger}\hat{b}\rangle\\
 & = & \sum_{r,k=0}c_{k}^{*}c_{r}e^{i\left(\Omega(k)-\Omega(r)\right)t/\hbar}r\langle N-k|N-r\rangle_{a}\langle k|r\rangle_{b}\\
 & = & \sum_{r=0}|c_{r}|^{2}r
\end{eqnarray*}

\begin{eqnarray*}
\langle\hat{a}^{\dagger}\hat{a}\hat{b}^{\dagger}\hat{b}\rangle & = & \sum_{r=0}\sum_{k=0}c_{k}^{*}c_{r}e^{i\left(\Omega(k)-\Omega(r)\right)t/\hbar}(N-r)r\times\\
 &  & \langle(N-k|N-r\rangle_{a}\langle k|r\rangle_{b}\\
 & = & \sum_{r=0}c_{r}^{2}(N-r)r
\end{eqnarray*}

\begin{eqnarray*}
\langle\hat{a}^{\dagger2}\hat{a}\hat{b}\rangle & = & \sum_{r=0}\sum_{k=0}c_{k}^{*}c_{r}e^{i\left(\Omega(k)-\Omega(r)\right)t/\hbar}\left(N-r\right)\times\\
 &  & \sqrt{N-r+1}\sqrt{r}\langle N-k|N-r+1\rangle_{a}\langle k|r-1\rangle_{b}\\
 & = & \sum_{k=0}^{N-1}c_{k}^{*}c_{k+1}e^{i\left(\Omega(k)-\Omega(k+1)\right)t/\hbar}\left(N-k-1\right)C(k)
\end{eqnarray*}
Here we have defined $C(k)=\sqrt{\left(k+1\right)\left(N-k\right)}$.

\begin{eqnarray*}
\langle\hat{a}^{\dagger}\hat{a}^{2}\hat{b}^{\dagger}\rangle & = & \sum_{r=0}\sum_{k=0}c_{k}^{*}c_{r}e^{i\left(\Omega(k)-\Omega(r)\right)t/\hbar}\left(N-r-1\right)\times\\
 &  & \sqrt{N-r}\sqrt{r+1}\langle(N-k|N-r-1\rangle_{a}\langle k|r+1\rangle_{b}\\
 & = & \sum_{r=0}^{N-1}c_{r+1}^{*}c_{r}e^{i\left(\Omega(r+1)-\Omega(r)\right)t/\hbar}\left(N-r-1\right)C(r)
\end{eqnarray*}
Here we have defined $C(r)=\sqrt{N-r}\sqrt{r+1}$.

\begin{eqnarray*}
\langle\hat{b}^{\dagger}\hat{b}\hat{a}\hat{b}^{\dagger}\rangle & = & \sum_{r=0}\sum_{k=0}c_{k}^{*}c_{r}e^{i\left(\Omega(k)-\Omega(r)\right)t/\hbar}\left(r+1\right)\times\\
 &  & \sqrt{N-r}\sqrt{r+1}\langle(N-k|N-r-1\rangle_{a}\langle k|r+1\rangle_{b}\\
 & = & \sum_{r=0}^{N-1}c_{r+1}^{*}c_{r}e^{i\left(\Omega(r+1)-\Omega(r)\right)t/\hbar}\left(r+1\right)C(r)
\end{eqnarray*}

\section{Analytical expression for the spin squeezing ratio\label{sec:ApB_AnExpSpinSqueezing-1}}

Here we show the analytical expressions in order to evaluate the spin
squeezing ratio given in Eq. (\ref{eq:spinsqparameterwine}) or $\bar{\xi}_{\theta}^{2}=\frac{\left\langle N\right\rangle \left(\Delta S_{\theta}\right)^{2}}{\left\langle S_{x}\right\rangle ^{2}}$.
Here we used the expression given in Appendix \ref{sec:ApAEvMoments-1}:
\begin{eqnarray*}
\langle\hat{S}_{x}\rangle & = & \frac{1}{2}(\langle\hat{a}^{\dagger}\hat{b}\rangle+\langle\hat{a}\hat{b}^{\dagger}\rangle)\\
 & = & \sum_{r=0}^{N-1}C_{n}C_{n+1}C(r)\cos\left[2(1-K)(N-2r-1)t/\hbar\right]\\
 & = & \sum_{r=0}^{N-1}\frac{N!}{2^{N}}\frac{e^{i2\chi(K-1)(N-1-2r)t/\hbar}}{r!(N-r-1)!}\times\\
 &  & \cos\left[2(1-K)(N-2r-1)t/\hbar\right]\\
 & = & \frac{Ne^{-2it(K(N-1)-N-1)}}{2^{N}\left(e^{4iKt}+e^{4it}\right)}\left(e^{4i(K-1)t}+1\right)^{N}
\end{eqnarray*}
Here we have used the definition of $\Omega(r)$, $C_{r}$ and $C(r)=\sqrt{N-r}\sqrt{r+1}$,
as well as 
\begin{eqnarray*}
\langle\hat{a}^{\dagger}\hat{b}\rangle & = & \sum_{r=0}^{N-1}C_{r}^{*}C_{r+1}e^{i\left[\Omega(r)-\Omega(r+1)\right]t/\hbar}\sqrt{(r+1)(N-r)}
\end{eqnarray*}

\[
\Omega(r)-\Omega(r+1)=2(1-K)(N-2r-1)
\]
Next, $(\Delta\hat{S}_{\theta})^{2}$ for the optimal angle is $(\Delta\hat{S}_{\theta})^{2}=\langle\hat{S}_{min}\rangle.$
From Eq. (\ref{eq:mins}) we get: 
\[
\langle\hat{S}_{\theta}^{2}\rangle_{min}=\frac{1}{2}(\langle\hat{S}_{y}^{2}\rangle+\langle\hat{S}_{z}^{2}\rangle)-\frac{\sqrt{4C^{2}+\vert F\vert^{2}}}{4}
\]
Similar to the case of $\langle\hat{S}_{x}\rangle,$ we use the definition
of $\Omega(r)$ and $C_{r}$ as well as the evaluation of the moments
given in Appendix \ref{sec:ApAdepth}. On simplifying terms we find:
\begin{eqnarray*}
\langle\hat{S}_{y}^{2}\rangle & = & -\frac{1}{4}\langle\hat{a}^{\dagger}\hat{a}^{\dagger}\hat{b}\hat{b}-\hat{a}^{\dagger}\hat{a}\hat{b}\hat{b}^{\dagger}-\hat{a}\hat{a}^{\dagger}\hat{b}^{\dagger}\hat{b}+\hat{a}\hat{a}\hat{b}^{\dagger}\hat{b}^{\dagger}\rangle\\
 & = & \frac{1}{8}\{N^{2}+N-4\sum_{r=0}^{N-2}\frac{2^{-N}N!}{r!(N-r-2)!}\times\\
 &  & \exp\left[-it4(K-1)(N-2(r+1))\right]\}\\
 & = & \frac{1}{8}\{N^{2}+N-\\
 &  & \frac{(N-1)Ne^{4it(-K(N-2)+N+2)}\left(1+e^{8i(K-1)t}\right)^{N}}{2^{N-2}\left(e^{8iKt}+e^{8it}\right)^{2}}\}
\end{eqnarray*}
\begin{eqnarray*}
\langle\hat{S}_{z}^{2}\rangle & = & \frac{1}{4}\langle(\hat{a}^{\dagger}\hat{a}-\hat{b}^{\dagger}\hat{b})\rangle=\frac{N}{4}
\end{eqnarray*}
\begin{eqnarray*}
 &  & F=\langle\hat{a}^{\dagger2}\hat{a}\hat{b}-\hat{a}\hat{b}^{\dagger}-\hat{a}^{\dagger}\hat{a}^{2}\hat{b}^{\dagger}+\hat{b}^{\dagger}\hat{b}(\hat{a}\hat{b}^{\dagger}-\hat{a}^{\dagger}\hat{b})\rangle\\
\\
 &  & =\frac{2^{-N}(N-1)N\left(e^{4it}-e^{4iKt}\right)e^{-2i(K+1)(N-1)t}}{\left(e^{4iKt}+e^{4it}\right)^{2}}\times\\
 &  & \left(e^{4iKNt}\left(1+e^{4i(1-K)t}\right)^{N}+e^{4iNt}\left(1+e^{4i(K-1)t}\right)^{N}\right)
\end{eqnarray*}
\begin{eqnarray*}
C & = & \langle\hat{S}_{z}^{2}\rangle-\langle\hat{S}_{y}^{2}\rangle\\
 & = & \frac{1}{8}[-N^{2}+N+\\
 &  & \frac{(N-1)Ne^{4it(-K(N-2)+N+2)}\left(1+e^{8i(K-1)t}\right)^{N}}{2^{N-2}\left(e^{8iKt}+e^{8it}\right)^{2}}]
\end{eqnarray*}
If we consider the case where $K=-1$ and $\chi=1$, the above terms
can be simplified: 
\begin{eqnarray*}
\langle\hat{S}_{x}\rangle & = & (N/2)\cos^{N-1}(4t).\\
\langle\hat{S}_{y}^{2}\rangle & = & \frac{1}{8}\left[N^{2}+N-(N-1)N\left(\cos^{N-2}(8t)\right)\right]\\
F & = & iN(N-1)\left(\cos^{N-2}(4t)\right)\sin(4t)\\
C & = & -\frac{1}{8}\left[N^{2}-N-(N-1)N\left(\cos^{N-2}(8t)\right)\right]
\end{eqnarray*}

\section{Discussion of mode entanglement}

Entanglement is a feature applying to quantum systems which are composites
of two or more physically distinguishable sub-systems. Both the overall
system and its sub-systems can be prepared in physically distinct
quantum states. For two sub-systems $A,B$, typical pure states for
these sub-systems can be listed as $\left\vert A\right\rangle ,\left\vert B\right\rangle $.
Pure states of the overall system are separable if they can be written
as $\left\vert A\right\rangle \otimes\left\vert B\right\rangle $,
otherwise they are entangled. Hence in general $(\left\vert A_{1}\right\rangle \otimes\left\vert B_{1}\right\rangle +$
$\left\vert A_{2}\right\rangle \otimes\left\vert B_{2}\right\rangle )$
is an entangled state. The definition can be generalised to mixed
states, where $\widehat{\rho}_{A},\widehat{\rho}_{B}$ are typical
sub-system mixed states. Mixed states are separable if they can be
written as $\widehat{\rho}_{A}\otimes\widehat{\rho}_{B}$, otherwise
they are entangled. Hence in general $\widehat{\rho}_{A_{1}}\otimes\widehat{\rho}_{B_{1}}+\widehat{\rho}_{A_{2}}\otimes\widehat{\rho}_{B_{2}}$
is an entangled state. We have ignored normalisation. Each sub-system
will also be associated with Hermitian operators $\widehat{\Omega}_{A},\widehat{\Omega}_{B}$
representing physical observables for the sub-system, and there will
also be Hermitian operators $\widehat{\Omega}$ involving operators
from both sub-systems (such as $\widehat{\Omega}_{A}\otimes\widehat{1}_{B}+\widehat{1}_{A}\otimes\widehat{\Omega}_{B})$
that will represent physical observables for the combined system.

In order to define entanglement in many particle systems three issues
arise - (1) How do we distinguish sub-systems from each other? (2)
Are there requirements that the states and observables for the sub-systems
and for the combined system must comply with? (3) How are cases where
the number of particles is not definite to be treated? In regard to
the third question, experimental situations do in fact arise (such
as in BECs) where particle numbers are not well-defined. The second
question is underpinned by the requirement that the states for the
sub-systems must be physically preparable and the observables physically
measureable. The first question reflects the idea that in regard to
sub-systems we are referring to an entity which has its own set of
physically preparable quantum states and observable quantities, and
which can exist independently without reference to other sub-systems.
It is particularly important to be precise about what sub-systems
are being referred to when discussing entanglement. A quantum state
which is entangled when referring to one choice of sub-systems may
well be separable when another choice is made - an example is given
below. With regard to these questions, there are two extreme situations
that could be involved. In the first situation the overall system
contains particles that are all identical. In the second situation
the overall system contains particles that are all different.

To treat systems of different particles the standard approach is to
use the first quantization formalism. Each distinct particle is associated
with a set of orthogonal one particle states (or modes) that it can
occupy. Note that the choice of modes is not unique - original sets
of orthogonal one particle states (modes) may be replaced by other
orthogonal sets. However, the single particle states for different
particles are obviously distinct from each other. Modes can often
be categorized as localized modes, where the corresponding single
particle wavefunction is confined to a restricted spatial region,
or may be categorized as delocalized modes, where the opposite applies.
Single particle harmonic oscillator states are an example of localized
modes, momentum states are an example of delocalized modes. Basis
states for the overall system or for sub-systems can be obtained as
products of the single particle states for each of the different particles
involved. Subject to certain restrictions discussed below, general
states are quantum superpositions of the basis states. These can represent
physically preparable states for either the overall system or for
a particular sub-system. Symmetrization principles applying for systems
of identical particles are irrelevant, and physical quantities for
each sub-system would be based on operators specific to the particles
involved (such as the momentum being the sum of momentum operators
for each particle), and hence being symmetrical under particle interchange
does not apply (Issue 2). Sub-systems are distinguished from each
other by just specifying which of the different particles they contain
(Issue 1), so sub-systems are defined by particles. Each sub-system
would therefore contain just one particle of each of the type involved.
Cases where the number of particles differ would be regarded as different
systems and each would have its own set of states. Compliance with
super-selection rules such as forbidding quantum states that involve
coherent superpositions of quantum states for different particles
(for example a linear combination of a neutron state with a proton
state) can be achieved by simply excluding such states as being unphysical
(Issue 2). Although the system is defined by the distinct particles
it contains, cases where the number of each particle is not definite
can be described via density operators involving statistical mixtures
of states with each having precise numbers ($0$ or $1$) of particles
of each type (Issue 3).

To treat systems of identical particles it is convenient to use the
second quantization formalism. The system is regarded as a quantum
field, which is associated with a collection of single particle states
(or modes). Again, the choice of modes is not unique - original sets
of orthogonal one particle states (modes) may be replaced by other
orthogonal sets, and modes may be localised or delocalised. The key
requirement is that the modes must be distinguishable from one another,
and this enables both the overall system and its sub-systems to be
specified via the modes that are involved - hence sub-systems can
be distinguished from each other (Issue 1). In this approach, particles
are associated with the occupancies of the various modes, so that
situations with differing numbers of particles will be treated as
differing quantum states of the same system, not as different systems.
In second quantization, Fock states defined via the occupancies of
the various modes are obtained from the vacuum state (containing no
particles in any mode) via the operation of mode creation operators,
and such states act as basis states for the quantum system or sub-system
being considered. These can represent physically preparable states
for either the overall system or a sub-system, and allowed general
states are quantum superpositions of the basis states. In linking
second and first quantization, the basis states are defined to be
in one-one correspondence with the symmetrized products of one particle
states that act as the basis states in the first quantization approach.
Creation and annihilation operators for each mode are defined to link
basis states where the occupancy changes by $\pm1$. The commutation
(anti-commutation) properties of the mode creation and annihilation
operators for bosons (fermions) reflect the first quantization requirement
that allowed physical states for these systems (and sub-systems) must
be symmetric (anti-symmetric) under the interchange of identical particles.
Furthermore, physical quantities in the first quantization approach
that satisfy the requirement of being symmetric under interchange
of identical particles are matched in second quantization by operators
based on mode annihilation and creation operators that are constructed
to have the same effect on the basis states (Issue 2). Compliance
with super-selection rules such as forbidding quantum states that
involve coherent superpositions of quantum states with differing numbers
of identical particles (for example a linear conbination of a one
boson state with a two boson state) can be achieved by simply excluding
such states on physical grounds (Issue 2). As the system is defined
by the distinct modes it contains, the case where the number of each
particle is not definite can be described via density operators involving
statistical mixtures of states with each having precise numbers (not
restricted to $0$ or $1$, apart from the case of a single mode system
for fermions) of particles (Issue 3).

Clearly, for identical particles an approach in which sub-systems
are specified by which modes are involved and which is based on using
the second quantization formalism is quite suitable for discussing
entanglement in such systems, since all three issues are resolved.
For distinguishable particles we can treat entanglement using an approach
in which sub-systems are specified by which particles are involved
and which is based on using the first quantization formalism. For
this case the introduction of the second quantization approach would
be superfluous. However, because each of the different particles is
associated with its own set of single particle states, it follows
that defining sub-systems via which particles they contain is actually
equivalent to defining them by which modes they contain - so in the
distinguishable particles case the particle approach is also equivalent
to the mode approach. However, the converse question is \textendash{}
Could the particle approach for defining sub-systems be applied in
the identical particles case based on the first quantization formalism?
The first problem is that there is no physical method that enables
us to distinguish one identical particle from another. In the first
quantization formalism we do label each identical particle with a
number, but when we then construct basis states with various numbers
of particles in the different one particle states, a symmetrization
operation is applied that treats them all the same. Similarly, all
the physical quantities are based on expressions in which each labelled
identical particle is included in the same way. Hence, if sub-systems
are defined by which labelled identical particles they contain, then
there is an immediate conflict with the requirement of being physically
distinguishable from another sub-system which has the same number
of differently labelled identical particles. There are of course no
numerical labels physically attached to each identical particles -
this is just a mathematical fiction. As a result of not being based
on distinguishable sub-systems, the labelled identical particle based
specification of sub-systems leads to states described in first quantization
being regarded as being entangled, whilst exactly the same state described
in second quantization (with sub-systems specified by modes) would
be regarded as separable. A simple illustration of this contradiction
occurs for a system of two bosons, in which one boson occupies a single
particle state $\left\vert \phi_{A}\right\rangle $ and the other
occupies a different single particle state $\left\vert \phi_{B}\right\rangle $.
With mode creation operators $\widehat{c}_{A}^{\dag}$ and $\widehat{c}_{B}^{\dag}$,\
in second quantization the quantum state is given by $\left\vert \Psi\right\rangle =\widehat{c}_{A}^{\dag}$
$\left\vert 0\right\rangle _{A}\otimes\widehat{c}_{B}^{\dag}$ $\left\vert 0\right\rangle _{B}=\left\vert 1\right\rangle _{A}\otimes\left\vert 1\right\rangle _{B}$,
which is a separable state for the combined system consisting of sub-systems
specified as modes $\phi_{A}$ and $\phi_{B}$. In first quantization
the same state is $\left\vert \Psi\right\rangle =\left(\left\vert \phi_{A}(1)\right\rangle \otimes\left\vert \phi_{B}(2)\right\rangle +\left\vert \phi_{B}(1)\right\rangle \otimes\left\vert \phi_{A}(2)\right\rangle \right)/\sqrt{2}$,
which would be regarded as an entangled state for the combined system
consisting of sub-systems specified by labelled identical particles
$1$ and $2$. As the labelled identical particle based specification
of sub-systems is in conflict with requirement for sub-systems to
be physically distinguishable, we believe that the mode based specification
of sub-systems is the correct one to apply in the case of systems
consisting of identical particles, and hence it is the approach used
in the present paper.

Some confusion can occur when discussing the effect of mode couplers
such as beam splitters on a quantum state. In general, the new state
may have a different entanglement status for the same pair of sub-system
to that of the original state. For the state $\left\vert \Psi\right\rangle $
above, it is well-known that the effect of a suitable beam splitter
could be described by a unitary operator $\widehat{U}$ such that
$\widehat{U}\widehat{c}_{A}^{\dag}\widehat{U}^{-1}=(\widehat{c}_{A}^{\dag}+\widehat{c}_{B}^{\dag})/\sqrt{2}$,
$\widehat{U}\widehat{c}_{B}^{\dag}\widehat{U}^{-1}=(-\widehat{c}_{A}^{\dag}+\widehat{c}_{B}^{\dag})/\sqrt{2}$.
In this case $\widehat{U}$ $\left\vert \Psi\right\rangle =\left(-\left\vert 2\right\rangle _{A}\otimes\left\vert 0\right\rangle _{B}+\left\vert 0\right\rangle _{A}\otimes\left\vert 2\right\rangle _{B}\right)/\sqrt{2}$,
which is now an entangled state for sub-systems specified as modes
$\phi_{A}$ and $\phi_{B}$. Thus in general, mode coupling creates
entanglement. For a different separable state given by $\left\vert \Phi\right\rangle =(1/\sqrt{N!})(\widehat{c}_{A}^{\dag})^{N}$
$\left\vert 0\right\rangle _{A}\otimes\left\vert 0\right\rangle _{B}=\left\vert N\right\rangle _{A}\otimes\left\vert 0\right\rangle _{B}$
the new state for the same beam splitter would be $\widehat{U}\left\vert \Phi\right\rangle =(1/\sqrt{N!})((\widehat{c}_{A}^{\dag}+\widehat{c}_{B}^{\dag})/\sqrt{2})^{N}(\left\vert 0\right\rangle _{A}\otimes\left\vert 0\right\rangle _{B})$,
and again the new state is an entangled state for sub-systems specified
as modes $\phi_{A}$ and $\phi_{B}$. However, if we introduce two
new orthogonal modes defined by the one particle states $\left\vert \phi_{C}\right\rangle =(\left\vert \phi_{A}\right\rangle +\left\vert \phi_{B}\right\rangle )/\sqrt{2}$
and $\left\vert \phi_{D}\right\rangle =(-\left\vert \phi_{A}\right\rangle +\left\vert \phi_{B}\right\rangle )/\sqrt{2}$,
then we see that the new state is also a separable state for sub-systems
specified as modes $\phi_{C}$ and $\phi_{D}$, having $N$ bosons
in sub-system $\phi_{C}$ and none in sub-system $\phi_{D}$. This
is a clear example of a quantum state that is entangled for one choice
of sub-systems yet is separable for another choice.

There is however, one situation for a system of identical particles
where the particle approach for defining sub-systems is appropriate.
This is where the sub-systems each consist of one or more localised
modes and the only states considered are where each sub-system just
contains one particle. Here each particle may be considered as distinguishable
from another one because it is just associated with distinguishable
localised modes. This situation applies for certain experiments in
quantum information theory, such as where two state identical qubits
each involving a single atom are localised by trapping in different
places. Reseachers in such quantum information situations generally
think of entanglement in terms of separated qubit sub-systems. However,
researchers in cold quantum gases are involved with identical particles
occupying delocalised modes, so here entanglements is best defined
in terms of modal sub-systems.

\end{document}